\documentclass[conference]{IEEEtran}
\ifCLASSINFOpdf
   \usepackage[pdftex]{graphicx}
\else
\fi
\usepackage{algorithmic}
\usepackage{url}
\usepackage{subfigure}
\usepackage{amsmath}
\usepackage[usenames,dvipsnames]{color}
\usepackage{ifthen}
\usepackage{comment}
\usepackage{epsfig}
\usepackage{amssymb}
\usepackage{amsfonts}

\makeatletter

\begin{document}
%
% paper title
% can use linebreaks \\ within to get better formatting as desired
\title{Optimizing Energy Storage Participation \\ in Emerging Power Markets}

% author names and affiliations
% use a multiple column layout for up to three different
% affiliations
%\author{}

% conference papers do not typically use \thanks and this command
% is locked out in conference mode. If really needed, such as for
% the acknowledgment of grants, issue a \IEEEoverridecommandlockouts
% after \documentclass

% for over three affiliations, or if they all won't fit within the width
% of the page, use this alternative format:
%
%\author{\IEEEauthorblockN{Michael Shell\IEEEauthorrefmark{1},
%Homer Simpson\IEEEauthorrefmark{2},
%James Kirk\IEEEauthorrefmark{3},
%Montgomery Scott\IEEEauthorrefmark{3} and
%Eldon Tyrell\IEEEauthorrefmark{4}}
%\IEEEauthorblockA{\IEEEauthorrefmark{1}School of Electrical and Computer Engineering\\
%Georgia Institute of Technology,
%Atlanta, Georgia 30332--0250\\ Email: see http://www.michaelshell.org/contact.html}
%\IEEEauthorblockA{\IEEEauthorrefmark{2}Twentieth Century Fox, Springfield, USA\\
%Email: homer@thesimpsons.com}
%\IEEEauthorblockA{\IEEEauthorrefmark{3}Starfleet Academy, San Francisco, California 96678-2391\\
%Telephone: (800) 555--1212, Fax: (888) 555--1212}
%\IEEEauthorblockA{\IEEEauthorrefmark{4}Tyrell Inc., 123 Replicant Street, Los Angeles, California 90210--4321}}

\author{\IEEEauthorblockN{Hao Chen\IEEEauthorrefmark{1},
Zhenhua Liu\IEEEauthorrefmark{2},
Ayse K. Coskun\IEEEauthorrefmark{1} and
Adam Wierman\IEEEauthorrefmark{3}}
\IEEEauthorblockA{\IEEEauthorrefmark{1}Electrical and Computer Engineering, Boston University}
\IEEEauthorblockA{\IEEEauthorrefmark{2}Applied Mathematics and Statistics, Stony Brook University}
%\IEEEauthorblockA{\IEEEauthorrefmark{3}Electrical and Computer Engineering
%Boston University}
\IEEEauthorblockA{\IEEEauthorrefmark{3}Computing and Mathematical Sciences, California Institute of Technology}}

% use for special paper notices
%\IEEEspecialpapernotice{(Invited Paper)}

% make the title area
\maketitle

\begin{abstract}
%\boldmath
The growing amount of intermittent renewables in power generation creates challenges for real-time matching of supply and demand in the power grid. Emerging ancillary power markets provide new incentives to consumers (e.g., electrical vehicles, data centers, and others) to perform demand response to help stabilize the electricity grid. A promising class of potential demand response providers includes energy storage systems (ESSs). This paper evaluates the benefits of using various types of novel ESS technologies for a variety of emerging smart grid demand response programs, such as regulation services reserves (RSRs), contingency reserves, and peak shaving. We model, formulate and solve optimization problems to maximize the net profit of ESSs in providing each demand response. Our solution selects the optimal power and energy capacities of the ESS, determines the optimal reserve value to provide as well as the ESS real-time operational policy for program participation. Our results highlight that applying ultra-capacitors and flywheels in RSR has the potential to be up to 30 times more profitable than using common battery technologies such as LI and LA batteries for peak shaving.
\end{abstract}
% IEEEtran.cls defaults to using nonbold math in the Abstract.
% This preserves the distinction between vectors and scalars. However,
% if the conference you are submitting to favors bold math in the abstract,
% then you can use LaTeX's standard command \boldmath at the very start
% of the abstract to achieve this. Many IEEE journals/conferences frown on
% math in the abstract anyway.

% no keywords

% For peer review papers, you can put extra information on the cover
% page as needed:
% \ifCLASSOPTIONpeerreview
% \begin{center} \bfseries EDICS Category: 3-BBND \end{center}
% \fi
%
% For peerreview papers, this IEEEtran command inserts a page break and
% creates the second title. It will be ignored for other modes.
\IEEEpeerreviewmaketitle

% \vspace{-0.1in}
\section{Introduction}
\label{sec: intro}
%\vspace{-0.05in}

A sustainable energy future mandates integrating a larger portion of renewable generation into the grid. Most states in the US and several European countries already have aggressive targets to increase the share of renewables in their portfolios~\cite{State, EU}. The fact that many forms of renewable generation are intermittent by nature (e.g., wind and solar) creates significant challenges for grid operators, who need to match supply and demand in real-time. In response to this challenge, emerging ancillary power markets provide sizable monetary incentives for the consumers to perform {\em demand response}, which refers to a consumer adjusting its own electricity use following a set of constraints or directives given by the grid operator.

Among potential demand response program participants, data centers, electrical vehicles (EVs), and smart buildings are especially promising, and have received recent attention from the research community~\cite{liu2014pricing, chenIGCC, brocanelli2013joint, wei2014co}. This attention is due to their significant flexibility in energy consumption, as well as the large cumulative power consumption levels and/or fast growth these entities provide.

One of the most promising participation opportunities for demand response comes from using energy storage systems (ESSs), which can potentially charge/discharge depending on the demand response program requirements reliably. There are a variety of energy storage startup companies~\cite{EnerNOC,comverge} that use ESSs to participate directly in energy market programs this way.  Additionally, entities such as data centers and smart buildings, which have on-site ESSs to manage power outages, can make use of ESSs to receive monetary incentives without having to alter their internal performance. ESSs have been studied for participation in well-known power programs such as real-time pricing~\cite{wang2013optimal}, but the potential of ESS participation in many of the most promising demand response programs has yet to be understood, including regulation service reserves (RSR), contingency reserves in emerging ancillary service markets, and peak shaving programs. Some recent work has begun to investigate these programs or ESS capacity planning~\cite{walawalkar2007economics, kumaraswamy2013evaluating, vu2009benefits, 7007628}; however, in most cases, these papers use simplified participation models, e.g., an RSR model that ignores regulation accuracy constraints and penalties~\cite{YoungIGCC14}. Besides, few work studies the decisions of reserve value and the ESS capacity planning. Furthermore, different ESS technologies (e.g., lead-acid (LA) batteries, lithium-ion (LI) batteries, ultra/super-capacitors (UC), flywheels (FW), and compressed air energy storage (CAES))  have contrasting properties, which can dramatically impact profits of participation in such programs. Systematic evaluation and comparison of the benefits of using these ESS technologies in a variety of demand response opportunities do not exist in current literature.

This paper's goal is to thoroughly evaluate, optimize, and contrast a range of ESS technologies for participation in a variety of promising demand response programs. Our method seeks to provide a strategy for the selection and management of ESSs for a broad range of consumers (data centers, EVs, smart buildings) to maximize the incentives received from ancillary power markets, and hence, to minimize the electricity cost while helping stabilize the grid. Our specific contributions are:

First, we provide detailed models and optimization solutions for participation of ESSs in multiple smart grid programs, including RSR, contingency reserves and peak shaving (Sections~\ref{sec: generalF} and~\ref{sec: prog}). We also design practical heuristic solutions that handle the real life probabilistic constraints in RSR provisioning problem (Section~\ref{sec: heuristicRS}). In each model, the cost of ESS equipment, the revenue received for demand response, and constraints required by the demand response program are formulated. The net profits are optimized based on these models, and the corresponding optimal decisions of reserve value, ESS capacity planning and the operational policies are derived. 
 %Based on these models, we optimize the net profit of various ESSs in these programs. 
 The generality and wide applicability of the models and solutions distinguish this paper from previous work.

Second, the proposed models and optimal solutions enable, for the first time in the literature, a thorough comparison of the benefits of different ESSs for participation in demand response opportunities (Section~\ref{sec: prog}). We highlight the ESS technology that is the most appropriate for each power program (and vice-versa). 
%determine the most profitable program for each ESS type. 
Results show that UC is the most profitable ESS for RSR, while LI battery is the best choice for peak shaving. Also, we show that none of today's typical ESSs can earn positive net profits from providing contingency reserves.

Finally, in addition to evaluation with offline optimization solutions, this paper proposes heuristic practical online policies for provisioning with different types of ESSs in RSR program  (Section~\ref{sec: online}), which is the most profitable program among those studied. As opposed to the offline solution, our online solution does not require information of RSR signal in advance, and thus, is applicable for real-life use. The solution adaptively leverages the tolerable RSR signal tracking errors for pursing larger profits. Our solution is able to satisfy all constraints and thus guarantee the feasibility of the provision, while still achieving significant profits.

%The remainder of the paper is organized as follows. Section~\ref{sec: generalF} characterizes various types of ESSs and introduces the models we use for each ESS technology. Section~\ref{sec: prog} optimizes, evaluates and compares ESSs in multiple demand response programs. Section~\ref{sec: policy} studies both online and offline policies for RSR provisioning. Section~\ref{sec: related} contains a detailed review of the related work. Section~\ref{sec: conclusion} provides concluding remarks.

 %\vspace{-0.1in}
\section{Energy Storage Systems}
\label{sec: generalF}
% \vspace{-0.05in}
%startup companies in markets: ESS/EV/data center (ES on site)

%The growing penetration of intermittent renewable energy sources, such as wind and solar, adds considerable difficulty to the operation of the power grid due to the that that such generation sources cannot be dispatched on demand, as is done for conventional generation sources. Energy storage systems (ESSs) are recognized as crucial to easing this difficulty since they allow for smoothing of intermittent sources.

%For example, today ESSs can participate in a variety of demand response programs, e.g., regulation service reserves (RSR), contingency reserves, peak shaving. Participants in these programs include electric vehicles~\cite{EV-SCE,EV-PGE}, data centers, which have ESSs on site by default \cite{urgaonkar2011optimal}, and various energy storage startup companies ~\cite{EnerNOC,comverge}.

%Traditionally, large scale ESSs have been deemed too expensive for widespread use in power systems; however this is beginning to change.  In particular, 

Storage technologies are becoming more cost-effective and wide spread and, at the same time, more lucrative market participation opportunities are emerging. 
% However, to this point, it is difficult to understand which storage technologies are suited for which market opportunities, and how much profit can be gained through participation. This is because different energy storage technologies have very different capabilities and constraints. 
In this section, we provide an overview of some potential storage technologies and define a model that enables us to study the participation of each ESS in various market opportunities.

 %\vspace{-0.1in}
\vspace{-0.08in}
\subsection{Background on Energy Storage Systems}
% \vspace{-0.05in}

In this paper, we focus on five popular ESSs, namely, lead-acid (LA) batteries, lithium-ion (LI) batteries, ultra/super-capacitors (UC), flywheels (FW), and compressed air energy storage (CAES). In the following, we briefly highlight important characteristics of each. The interested reader can refer to prior work~\cite{PSUSigmetrics12} for more information.

\textbf{Lead-Acid (LA) batteries} are widely used in daily life, e.g., in car batteries. They have very low self-discharge loss rates, which makes them suitable for the demand response programs with long durations, e.g., hours. Additionally, they have moderate energy cost and power cost, and therefore are robust under different market scenarios. However, the key disadvantage of LA batteries is the relatively small number of charge/discharge cycles and shorter float life. LA batteries can only be used for several thousand circles. %This limits the feasibility of using LA batteries in demand response programs that require frequent charge/discharge, e.g., RSR.

\textbf{Lithium-Ion (LI) batteries} are also widely used in our daily life, and have similar characteristics to LA batteries.  The key difference is that LI batteries have relatively higher costs, longer lifetimes, more cycles, and higher efficiency.

\textbf{Ultra/super-Capacitors (UCs)} differ dramatically from LI and LA batteries. UCs have an extremely high tolerance for frequent charging/discharging. Additionally, UCs have high efficiency and power density. However, they have a high energy cost (around \$10,000/kWh) and high self-discharge rate. %Thus, one could expect UCs to be suitable for demand response programs such as RSR. %Furthermore, if ESS can be organized hierarchically similar to computer storage hierarchy, UC can be utilized to improve the overall

\textbf{Flywheels (FWs)} represent a middle ground between LI/LA batteries and UCs.  Like UCs, they have high efficiency and power density, but also high energy cost and a high self-discharge rate. %It can be considered as a tradeoff between normal batteries and UC. %\adam{need more info here}

\textbf{Compressed Air Energy Storage (CAES)} has a very low energy cost and self-discharge rate.  However, %it has a key weakness compared to the above technologies -- 
it has a very slow ramping time (10 min vs. 1ms in the other four ESSs).  This means that it cannot adapt quickly, which limits participation of CAES in some market programs.  Additionally, it has a very low energy density (large space needed) and a high power cost. %\adam{need more info here}

\begin{table}[tb]
\caption{A Selection of Today's Typical Capacities of ESSs, Based on Space Constraints.} 
\label{tb:typicalESD}
\begin{minipage}{8cm}
\def\arraystretch{1.5}\tabcolsep 0.8pt
\def\thefootnote{a}\footnotesize
 \centering
\begin{tabular}{|c|c|c|c|c|c|}
\hline
& LA & LI& UC & FW & CAES\\
\hline
$P_{cap}$ (kW) & 1,000 & 1,000 & 20,000  & 10,000 & 20\\
\hline
$ E_{cap}$ (kWh) & 250 & 250 & 250 & 250 & 250\\
\hline
\end{tabular}
\end{minipage}
\vspace{-0.12in}
\end{table}

 \vspace{-0.08in}
\subsection{Modeling Energy Storage Systems}
%  \vspace{-0.05in}
There are two key components in modeling ESSs: costs (both of procurement and operation) and operation constraints (self-discharging, ramping, etc.). Operation constraints can be classified into (i) constraints imposed by the ESS technology and (ii) constraints imposed by the demand response program. Constraints of type (i) are discussed here, and constraints of type (ii) are discussed in Section \ref{sec: prog}.

\textbf{ESS Costs:} The life span of an ESS is normally years with one-time upfront purchase/installation cost, yet participation in a demand response program can span a year, a month, or even a day. In order to handle the mismatch in time granularity, we amortize the upfront cost evenly
%\footnote{More advanced methods can be employed and our framework can be applied more broadly with some adaptations.} 
over the lifespan of the ESS.
Let $P_{cap}$ (in kW) and $E_{cap}$ (in kWh) represent the power capacity and energy capacity of the ESS, respectively, and $\Pi^P$ (in \$/kW) and $\Pi^E$ (in \$/kWh) are the corresponding prices. Then the one-time upfront cost is\footnote{Other ways of calculating the upfront cost exist (e.g., the upfront cost is selected as the maximum of the costs on power capacity and energy capacity~\cite{PSUSigmetrics12}). Our method is adaptable to such calculations, e.g., an ancillary variable can be introduced to convert the selection of the maximum on power and energy capacities into two linear constraints.
}:
\vspace{-0.05in}
%\begin{small}
\begin{equation}
\begin{aligned}
\Pi^P P_{cap} + \Pi^E E_{cap}.
\end{aligned}
\end{equation}
%\end{small}

\vspace{-0.05in}
Two factors that need to be considered to calculate the duration of use are the face-plate lifetime $T_{\max}$ and the maximal number of charge/discharge cycles $L_{cyc}$. Assuming the charge/discharge frequency is $f_j$, the effective duration of use is $\min \left\{T_{\max}, \frac{L_{cyc}}{f_j}\right\}$. Since many of the demand response programs clear the credits daily, we amortize the cost of ESSs into daily prices, namely, for each type of ESS $k$, we define its daily power and energy capacity prices as $\Pi^{P,d}_{k}$ and $\Pi^{E,d}_{k}$ as:
\vspace{-0.1in}
\begin{small}
\begin{equation}
\begin{aligned}
\label{gm: first}
\Pi^{P,d}_{k} = \frac {\Pi^P_k} { \min \left\{ T_{\max}, \frac{L_{cyc}}{ f_j}   \right\} }, \ \
\Pi^{E,d}_{k} = \frac {\Pi^E_k} { \min \left\{  T_{\max}, \frac{L_{cyc}}{f_j}  \right\} },
\end{aligned}
\end{equation}
\end{small}
where $f_j$ is the frequency of the charge/discharge in program $j$. Therefore, the daily amortized cost is:
\vspace{-0.05in}
%\begin{small}
\begin{equation}
\begin{aligned}
\label{gm: esscost}
\Pi^{P,d}_{k} P_{cap} + \Pi^{E,d}_{k} E_{cap}.
\end{aligned}
\end{equation}
%\end{small}

%The objective of the problem is to maximize the net profits received from applying ESSs in participating demand response and ancillary market programs. The net profits equal to the credits received from the demand response participation minus the cost of ESSs. Hence the objective function is:
%\begin{equation}
%\begin{aligned}
%\label{gm: first}
%\text{Maximize} \ \ Credits_j - ESSCost_k,
%\end{aligned}
%\end{equation}
%where $j$ and $k$ represent different types of programs and ESS equipments, respectively.

%The cost of ESSs is a one time cost charged when an equipment is bought. An ESS is charged based on its power capacity $P_{cap}$, and energy capacity $E_{cap}$. The corresponding prices of them, $\Pi^P$ and $\Pi^E$, vary with different types of ESSs. An ESS is typically designed of being used for several years, with a face-plate lifetime $T_{max}$. However, the real lifetime of the ESS also depends on how it is used. For example, the lifetime of the lead-acid (LA) battery or the lithium-ion (LI) battery can be greatly shortened if the battery is charged and discharged too frequently, or with a large Depth-of-Discharge (DoD). The lifecycle, $L_{cyc}$, i.e., the maximal number of charge/discharge cycle of an ESS, depends also on DoD. In this work, we simply use some typical values of DoD and $L_{cyc}$ for each type of ESSs reported from previous work~\cite{}, and do not study their relations.
\vspace{-0.05in}
\textbf{ESS Operation Constraints:} Assume that at time $t$, the charge and discharge rates of an ESS are $r_t$ and $d_t$, respectively. We denote the total energy stored in the ESS at time $t$ as $e_t$, and the overall power rate from the view of the system level as $u_t$. Then we have:
\vspace{-0.05in}
%\begin{small}
\begin{equation}
\begin{aligned}
\label{eq4}
e_{t}  =  e_{t-1} - \mu e_{t-1} + r_t - d_t, \  \forall t, \\
u_t = r_t / \eta - d_t, \  \forall t,
\end{aligned}
\end{equation}
%\end{small}
where $\mu$ is the self-discharge rate of the ESS, and $\eta$ is the energy charging efficiency. We have $\eta<1$, as there is always amount of loss during the ESS charge process. $\mu$ and $\eta$ vary with types of ESSs. For example, UC and FW in general have higher efficiency than LA and LI batteries, however, they have much higher self-discharge rate.

The charge and discharge rates are also constrained by the charge/discharge capacities of  the ESS, as follows:
\vspace{-0.05in}
%\begin{small}
\begin{equation}
\begin{aligned}
0 \leq r_t \leq  \frac{P_{cap}}{\gamma}, \  0 \leq d_t \leq P_{cap}, \ \forall t,
\end{aligned}
\end{equation}
%\end{small}
%\vspace{-0.05in}
where $P_{cap}$ is the power capacity of the ESS defined before, $\gamma$ is the ratio of discharge rate to charge rate. For UC and FW, $\gamma$ is close to 1, which means they have almost same charge and discharge capacities, however for LA and LI batteries, $\gamma>1$, representing a (much) slower recharge rate.

The amount of energy that is stored in the ESS is constrained by the ESS energy capacity $E_{cap}$. In addition, it is constrained by the Depth of Discharge (DoD), which helps guarantee the lifetime of the equipment:
%\begin{small}
\begin{equation}
\begin{aligned}
(1 -  DoD) E_{cap} \leq e_t \leq E_{cap}, \  \forall t.
\end{aligned}
\end{equation}
%\end{small}

\vspace{-0.1in}
Finally, though most ESSs are able to ramp up their discharge rate extremely fast, some ESSs, e.g., CAES, cannot. Thus, we have the discharge rate ramp up constraint:
\vspace{-0.05in}
%\begin{small}
\begin{equation}
\begin{aligned}
\label{gm: last}
d_{t+1} - d_{t} \leq \frac{P_{cap}}{T^{ramp}}, \ \forall t, \\
\end{aligned}
\end{equation}
%\end{small}
where $T^{ramp}$ is the time for ESS to ramp up the discharge rate from 0 to $P_{cap}$. %counted in the number of time slots.

 %\vspace{-0.1in}

\section{Market Opportunities for \\ Energy Storage Systems}
\label{sec: prog}
% \vspace{-0.05in}

In this section, we propose detailed models of ESS participation in various electricity market programs, including RSR, contingency reserves, and peak shaving. % Then we compute optimal solutions and evaluate the potential benefits of each type of ESS in participating these energy market opportunities. 
We introduce the revenue function, $Revenue_j$ that represents the revenue received from participation in the program $j$, and the constraints, $Constraint_j$ that are required by the program operator. The net profit of participation equals to $Revenue_j$ minus the daily amortized cost of ESS in Eq.\eqref{gm: esscost}. For each type of ESS $k$ and each program $j$, we derive the optimal selections of ESS energy and power capacities, as well as the optimal ESS operational policy (including the amount of reserves to provide, and the solution of how to dynamically charge and discharge over time, etc.) for maximizing profit. Then we evaluate applying these ESSs with today's typical capacities, and conduct sensitivity analysis of the maximal net profit on the price of reserves. Finally, we compare the benefits of  these ESSs participating in each program.

\vspace{-0.1in}
\begin{figure}[tb]
\centering
\vspace{-0.2in}
\subfigure[Profit of LI batteries (\$/day).]{
\includegraphics[width=.22\textwidth]{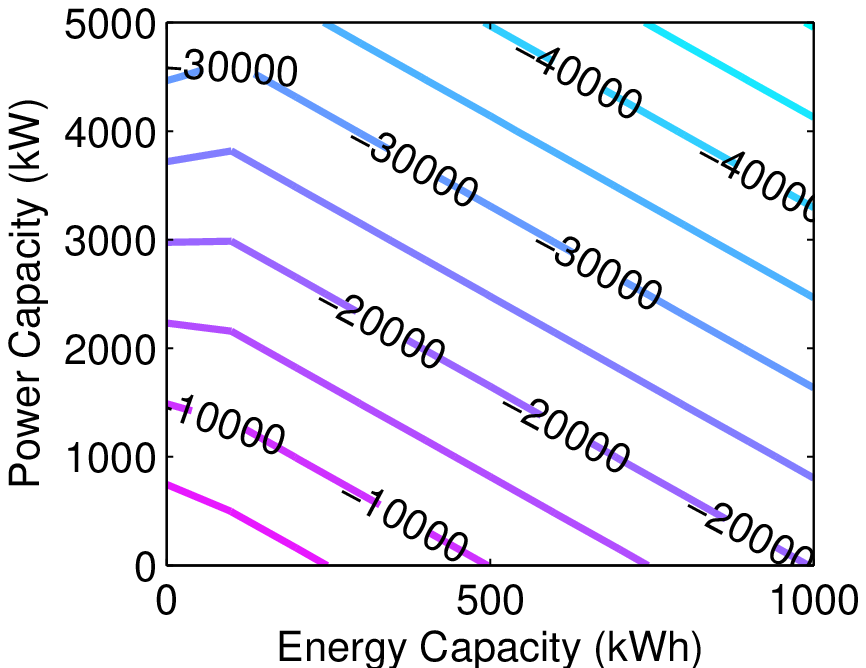}
 \label{fig:Li_SAonPE_fval}
}
\vspace{-0.1in}
\subfigure[Profit of UC (\$/day).]{
\includegraphics[width=.22\textwidth]{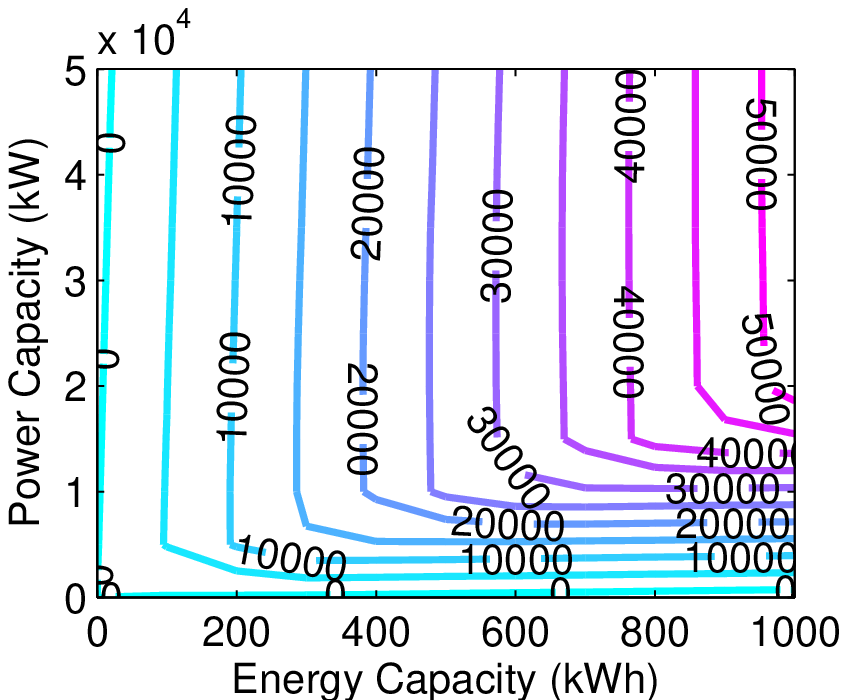}
 \label{fig:UC_SAonPE_fval}
}
\subfigure[Impacts of $R$ on net profit.]{
\includegraphics[width=.22\textwidth]{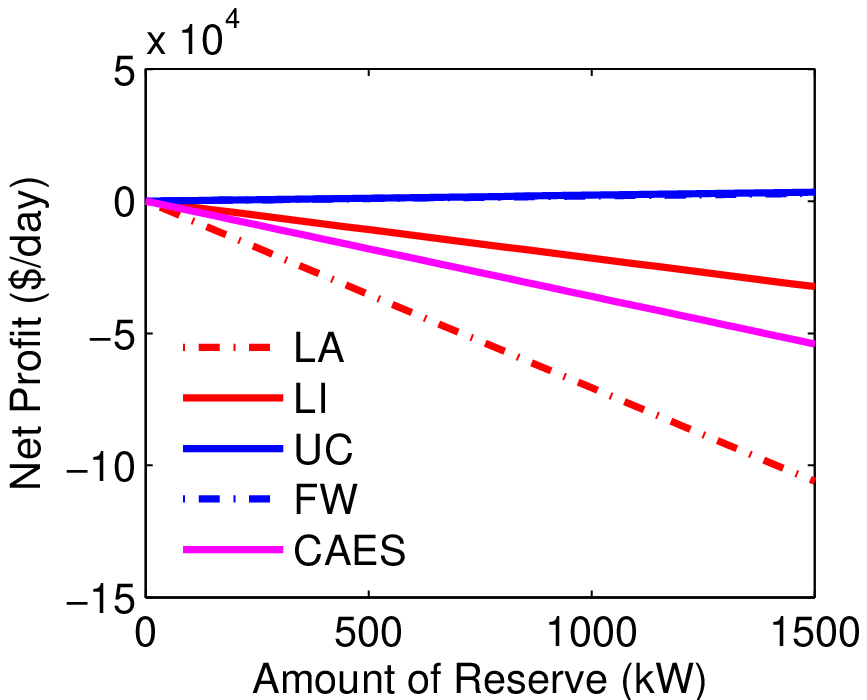}
\label{fig:SAonR_fval}
}
\vspace{-0.1in}
\subfigure[Impacts of price on net profit.]{
\includegraphics[width=.22\textwidth]{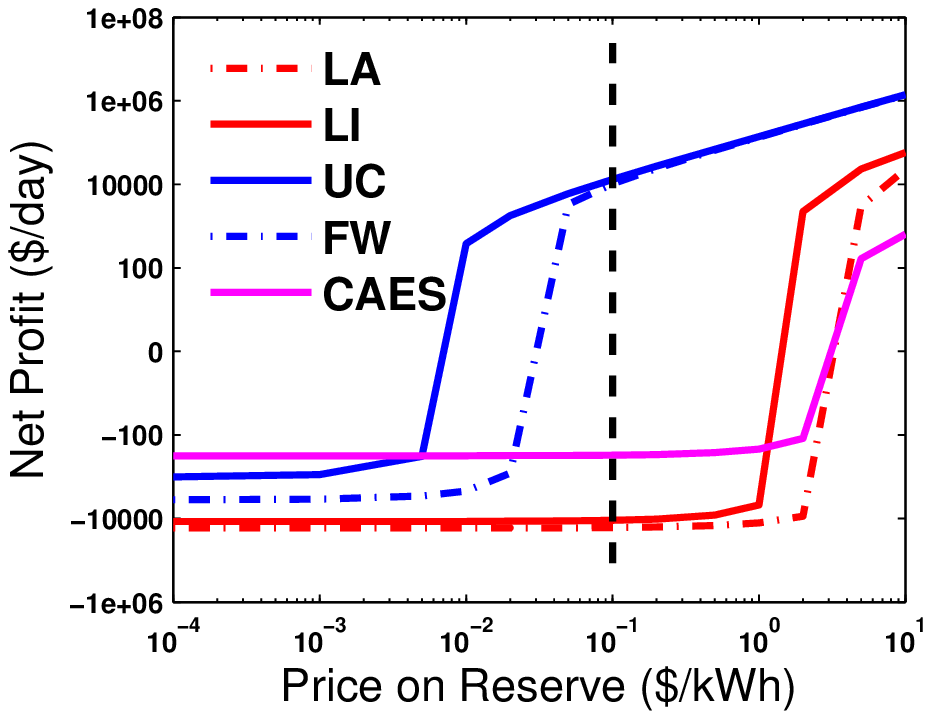}
\label{fig:RS_SAonRPrice}
}
%\subfigure[Operational policy (LI)]{
%\includegraphics[width=.22\textwidth]{figures/Li_track}
%\label{fig:Li_track}
%}
%\subfigure[Operational policy (UC)]{
%\includegraphics[width=.22\textwidth]{figures/UC_track}
%\label{fig:UC_track}
%}
%\vspace{-0.05in}
\caption{ESSs in regulation service reserves. %(a) and (b) show the optimal net profit via varying energy and power capacities for LI batteries and UC; (c) and (d) show the optimal net profit via varying amount of reserve provision, and the varying reserve price $\Pi^{RS}$, respectively for various ESSs. The black dashed line in (d) shows the current $\Pi^{RS}$; (e) and (f) show examples of the optimal operational policy (in an hour) in the case of $R= 550kW$ for LI batteries and UC.
}
\label{fig:RS}
 \vspace{-0.15in}
\end{figure}

% \vspace{-0.05in}
\subsection{Regulation Service Reserves (RSR)}
\label{sec: RSR}
%\vspace{-0.05in}

Historically, RSRs were mainly provided by centralized generators, but market rules are changing to encourage demand-side participation. This emerging demand response opportunity is quite attractive due to the high payments comparable to the real-time market price~\cite{pjmweb,aikema2012data}. RSR programs are typically quite demanding for participants.  Each RSR provider is obligated to modulate its power to track an RSR signal $\beta_t$ broadcast every 4 seconds (this defines the length of one time slot) by the independent system operator (ISO)~\cite{pjmweb}. 
%$\beta_t$ is in fact the output of a proportional-integral filter of system wide frequency deviation and Area Control Error (ACE), determined carefully in order to complement primary control reserves. 
The signal is between $[-1, 1]$, with an average of zero over long time intervals. It is updated every 4 seconds in increments that do not exceed $\pm 4/\tau$, where $\tau$ is in 100-300 seconds~\cite{chenASPDAC}.

% \vspace{-0.05in}
\subsubsection{Problem Formulation}

A provider receives $\Pi^{RS} \cdot R$ revenue for providing $R$ (kW) amount of reserves, where $\Pi^{RS}$ is the price of reserves. The revenue is reduced based on the tracking error of the RSR signal, i.e., $|u_t - R\beta_t|$, where $u_t$ is the power rate defined in Eq.\eqref{eq4}. The overall daily revenue received from RSR participation ($T =$ 1 day) is:
\vspace{-0.1in}
%\begin{small}
\begin{equation}
\begin{aligned}
\label{RS: credit}
Revenue_{RS} = \Pi^{RS} R -  \theta \cdot \Pi^{RS} (\frac{1}{T}\sum_{t=1}^T |u_t - R\beta_t|),
\end{aligned}
\vspace{-0.05in}
\end{equation}
%\end{small}
% \vspace{-0.05in}
where $\theta$ is the penalty coefficient on the tracking error.

The provider may lose the RSR contract if the constraint on signal tracking performance is violated. We formulate this using a probabilistic constraint:
\vspace{-0.1in}
 \begin{equation}
  \begin{aligned}
  \label{RS: cons}
\sum_{t=1}^T \mathbb{I}_{\{|\frac{u_t}{R\beta_t} - 1 | \leq \rho_1\}} \geq  \rho_{2}T\\
 \end{aligned}
  \end{equation}
where $\rho_1$ and $\rho_2$ are parameters set by the ISO. This equation shows that the probability of tracking error at each time $t$, (i.e., $|u_t - R\beta_t|$) that is smaller than $\rho_1R|\beta_t|$ should be greater than or equal to $\rho_2$. 

Putting Eq.\eqref{gm: first} - Eq.\eqref{RS: cons} together, the overall optimization formulation of ESSs in RSR is:
\vspace{-0.1in}
\begin{small}
 \begin{equation}
 \begin{aligned}
  \label{RS:whole}
{\underset{E_{cap}, P_{cap}, R,  \mathbf{r},  \mathbf{d},  \mathbf{u},  \mathbf{e}}{\max}} \ \ \  \Pi^{RS} R- \theta \cdot \Pi^{RS} \frac{1}{T}\sum_{t=1}^T |u_t - R\beta_t|  \\
 - (\Pi^{P,d} P_{cap} + \Pi^{E,d} E_{cap} ), \\
\textbf{s.t.} \ \ \ \sum_{t=1}^T \mathbb{I}_{\{|\frac{u_t}{R\beta_t} - 1 | \leq \rho_1\}} \geq  \rho_{2}T,\\
e_{t}  =  e_{t-1} - \mu e_{t-1} + r_{t} - d_{t}, \ \forall t \in [1, T],\\
u_t = r_t / \eta - d_t, \ \forall t \in [1, T] ,\\
0 \leq r_t \leq \frac{P_{cap}}{\gamma}, \ 0 \leq d_t \leq P_{cap},\ \forall t \in [1, T], \\
(1 -  DoD) E_{cap} \leq e_t \leq E_{cap}, \ \forall t \in [0, T], \\
d_{t+1} - d_{t} \leq \frac{P_{cap}}{T^{ramp}}, \ \forall t \in [1, T-1], \\
P_{cap} \geq 0, E_{cap} \geq 0, R \geq 0.\\
\end{aligned}
\end{equation}
\end{small}
In the formulation we use $\mathbf{r}$,  $\mathbf{d}$,  $\mathbf{u}$ and  $\mathbf{e}$ to denote the vectors of $r_t$, $d_t$, $u_t$ and $e_t$, respectively. The objective function is to maximize the net profit of the participation, recalling that the net profit equals the revenue for providing reserves (reduced by the tracking error) minus the amortized cost of ESS equipment. The constraints are imposed by both the demand response program (RSR here) and the ESS technology. The decision variables of this optimization problem are:
\vspace{-0.05in}
\begin{itemize}
\item Power and energy capacities of ESS, i.e., ($P_{cap}$, $E_{cap}$);
\vspace{-0.18in}
\item The amount of reserve to provide, i.e., $R$;
\vspace{-0.05in}
\item $\mathbf{r}$,  $\mathbf{d}$,  $\mathbf{u}$ and  $\mathbf{e}$, which represent how the ESS is operated dynamically, i.e., the operational policy.
\end{itemize}
%\vspace{-0.05in}
% The absolute value in above formulation leads to a piecewise linear property. We simplify the piecewise linear formulation to a linear one by introducing ancillary variables $z_t^+$ and $z_t^-$ satisfying:
% \vspace{-0.1in}
% \begin{equation}
%  \begin{aligned}
%  \label{eq: trans}
%|u_t - R\beta_t | = z_t^+ + z_t^- , \ \forall t \in [1, T], \\
%u_t - R\beta_t = z_t^+ - z_t^-, \ \forall t \in [1, T], \\
%z_t^+ \geq 0, \ z_t^- \geq 0 , \ \forall t \in [1, T]. \\
%  \end{aligned}
%  \end{equation}

\subsubsection{Case Study}

To evaluate the potential value from RSR program, we solve the above optimization formulation for the types of ESSs introduced before. We use parameters defined by prior work~\cite{PSUSigmetrics12}. The RSR signal $\beta_t$ that we use is a real 24-hour signal from PJM~\cite{pjmweb}. Additionally, $\rho_1=0.2$, $\theta=1$ and $\Pi^{RS} = \$0.1$/kWh based on today's markets~\cite{aikema2012data}.

The probabilistic constraint makes Eq.\eqref{RS:whole} not straightforward to solve. To simplify the problem, we first study the case of  $\rho_2=1$, in which the probabilistic constraint in Eq.\eqref{RS: cons} can be transformed to a deterministic constraint:
\vspace{-0.05in}
 \begin{equation}
  \begin{aligned}
  \label{eq: 100Cons}
\left|\frac{u_t}{R\beta_t} - 1 \right| \leq \rho_1, \forall t \in [1, T].\\
  \end{aligned}
  \end{equation}
Heuristic solutions of $\rho_2 < 1$ will be discussed in Section~\ref{sec: policy}. Finally, the absolute value on the tracking error in Eq.(10) and Eq.(11) leads to piecewise linear property. We simplify the piecewise linear formulation to a linear one by introducing ancillary variables.

%\textcolor{green}{Finally, the absolute value on the tracking error in Eq.\eqref{RS:whole} and Eq.\eqref{eq: 100Cons}  leads to piecewise linear property. We simplify the piecewise linear formulation to a linear one by introducing ancillary variables $z_t^+$ and $z_t^-$ satisfying:
%\vspace{-0.05in}
% \begin{equation}
 % \begin{aligned}
 % \label{eq: trans}
%|u_t - R\beta_t | = z_t^+ + z_t^- , \ \forall t \in [1, T], \\
%u_t - R\beta_t = z_t^+ - z_t^-, \ \forall t \in [1, T], \\
%z_t^+ \geq 0, \ z_t^- \geq 0 , \ \forall t \in [1, T]. \\
 % \end{aligned}
 % \end{equation}
%In this way, we convert Eq.\eqref{RS:whole} into a linear programming problem, and the optimal solution can be solved. }

At the current reserve prices ($\Pi^{RS} = \$0.1$/kWh), the optimal solution of Eq.\eqref{RS:whole} for LA, LI batteries and CAES are all $P^*_{cap} = E^*_{cap} = R^* = 0$, which demonstrates that there is no net profit of LA, LI batteries or CAES to participate in RSR program, i.e., the ESS cost of them is always larger than the revenue received from the program, no matter what the power and energy capacities are used or how they are operated dynamically. On the other hand, there is no feasible optimal solution of Eq.\eqref{RS:whole} for UC and FW: the net profit keeps increasing as $P_{cap}$, $E_{cap}$ and $R$ increase, which demonstrates that the maximal net profit is large for UC and FW, as long as sufficiently large power and energy capacities can be offered. This highlights that the revenue earned by UC and FW from RSR is always larger than the amortized cost of the ESS.   

% The simulation results demonstrate that at current reserve prices, the maximal net profit of LA, LI batteries and CAES are 0, i.e., the ESS cost of them is always larger than the revenue from the program, no matter what the power and energy capacities are used or how they are operated dynamically. On the other hand, the maximal net profit is large for UC and FW, as long as sufficiently large power and energy capacities can be offered. This highlights that the revenue earned by UC and FW from RSR is always larger than the amortized cost of the ESS.

We then study the sensitivity of net profit to energy, power capacities and the amount of reserve provision. Fig.\ref{fig:Li_SAonPE_fval} and Fig.\ref{fig:UC_SAonPE_fval} present the optimal net profit (the negative value represents that the cost of ESS is larger than the revenue, hence the net profit is less than 0) for varying energy and power capacities ($E_{cap}$, $P_{cap}$), and for LI batteries and UC respectively, in contour plots. LA batteries have similar results to LI batteries, and FW is similar to UC. From the figures, we see that for LA/LI batteries, the net profits of participating RSR are always negative, and the larger capacities of them are used, the higher cost there would be. On contrary, for UC and FW, a larger ($E_{cap}$, $P_{cap}$) creates larger net profit. The optimal net profit via varying amount of reserve, i.e., $R$, is shown in Fig.\ref{fig:SAonR_fval}. The net profits of LA, LI batteries and CAES are always negative and monotonously decrease along the increase of $R$, while the net profits of UC and FW are always larger than 0 and monotonously increase. Note that for all ESSs, providing larger $R$ requires larger ESS capacities.
%which further proves that there is no profit of LA, LI batteries and CAES to participate in RSR program at today's reserve prices, while the profit of participation by UC and FW can be huge. 
%\textcolor{green}{Fig.\ref{fig:Li_track} and Fig.\ref{fig:UC_track} demonstrate examples of the optimal dynamic operational policy (in an hour) in the case of $R$= 550kW for LI batteries and UC, respectively.}

% A number of interesting points can be seen from these figures.  For example, for all ESSs, providing larger $R$ requires larger capacities ($E_{cap}$, $P_{cap}$).  Additionally, for ESSs such as  LA/LI batteries, and CAES, providing larger $R$, or using larger ($E_{cap}$, $P_{cap}$) to participate RSR leads to larger cost.  Finally, for ESSs such as UC and FW, larger $R$ or ($E_{cap}$, $P_{cap}$) creates larger net profit.

The main factors that lead to such differences among ESSs are related to the characteristics of the ESSs. Since the RSR signal changes rapidly (every 4 seconds) and bidirectionally, in order to track it, RSR providers must have a large power capacity and large charge/discharge cycles. A large energy capacity, however, is not necessary, as the RSR signal has an average of zero over longer time intervals. UC and FW perfectly match these RSR characteristics: they have extremely high tolerance for frequent charging/discharging, high efficiency and power density, and relatively low power capacity cost, whereas under the high charge/discharge frequency in RSR, the lifetime of LA or LI batteries is shortened to less than 10 days due to the limited life cycle, which results in great cost and thus they no longer gain any net profit from RSR participation. CAES is even more limited due to the very large ramp up delay in discharge and the extremely small power density.

% In practice, the ESSs power and energy capacities have upper bound limitations.
Next we focus on the RSR participation of different ESS technologies with today's typical capacities.
%Ideally, one can buy capacities of ESSs as large as they want. However 
In practice, the power and energy capacities of ESSs usually have upper bound limitations due to the restrictions of manufacturing techniques, unit prices and space constraints. Table~\ref{tb:typicalESD} lists a selection of today's typical capacities of different types of ESSs referring to recent work~\cite{PSUSigmetrics12, mccluer2008comparing, SmithAdavance, ghiassi2013towards}, estimated mainly based on space constraints\footnote{Since we have taken the cost and unit price information into account in the problem formulation, we no longer consider it as a problem in determining typical capacities of ESSs here.}. The power capacity of CAES is small due to its extremely small power density. The optimal net profit and the corresponding optimal $R^*$ of these typical ESSs in RSR are listed in the $3^{rd}$ row of Table~\ref{tb:compare}\footnote{All results listed in Table~\ref{tb:compare} are the optimization solutions of Eq.\eqref{RS:whole} when $E_{cap}$ and $P_{cap}$ are given as in Table~\ref{tb:typicalESD}.}. From the table, today's typical UC or FW can provide around 6MW RSR, and gain more than \$10,000 net profit a day, which are close to the power consumption and the cost of a data center with 10,000-20,000 servers. The cost of this typical UC or FW is around \$4 million, which can be paid back in less than one year by receiving RSR credit.

%\textcolor{red}{Finally, we study the sensitivity of the net profit to the reserve price.} 
Fig.~\ref{fig:RS_SAonRPrice} shows the optimal net profit via varying reserve price $\Pi^{RS}$, for different types of ESSs with their capacities fixed and given in Table~\ref{tb:typicalESD}. The black dashed line represents where the current market reserve price is around. From the figure, LI, LA batteries and CAES start to gain net profit (the value of the net profit is larger than 0) when the reserve price $\Pi^{RS} $ is beyond \$1/kWh.

\vspace{-0.1in}
\subsection{Contingency Reserves}
\label{sec: CR}
% \vspace{-0.05in}

In ancillary markets, contingency reserves are used to respond to loss of power supplies during generation or line failures. They are typically called by the market less than once a day, and some of them are called even less than once a year. A call typically lasts from several minutes to a few hours. Reserves that are able to respond immediately are known as {\it spinning reserves}, whereas reserves that require more time to respond are called {\it non-spinning reserves}. For example, NYISO provides 10-minute spinning and 10-minute non-spinning reserves. Another type of reserves, the {\it operating reserves}, are also provided by NYISO, as supplements of other reserves. Operating reserves have longer reaction time but also last longer, e.g., more than 30 minutes~\cite{aikema2012data}. 10-minute spinning reserves have the highest price while the price of 30-minute operating reserves is the lowest. All these prices are significantly lower than that of RSR. Overall, due to the much lower frequency of calls as well as the lower price of the reserves, the revenue received from contingency reserve provision is much lower than revenue from RSR provision.

%Though {\it emergency demand response} generally does not belong to ancillary services, we introduce it here as it has very similar characteristics to contingency reserves. Emergence demand response is called when other reserve categories are likely to be inadequate in making up the loss of supplies. It is called less than once a year on average, with each time lasting for at least 4 hours. It does not react immediately, instead, it is mostly based on day-ahead requests~\cite{aikema2012data}. Though the price of emergency demand response can be as high as the real-time energy price, high credit is never expected from emergency demand response due to the extremely low frequency that it is called at.
%\vspace{-0.05in}
\subsubsection{Problem Formulation}

The revenue of contingency reserves can be modeled as:
\vspace{-0.05in}
\begin{equation}
\begin{aligned}
\label{CR: credit}
Revenue_{CR} = \Pi^{CR} R,
\end{aligned}
\vspace{-0.05in}
\end{equation}
where $R$ is the amount of contingency reserves provided and $\Pi^{CR}$ is the price of the reserve. Unlike RSR, the contingency reserve provision is single directional with:
\vspace{-0.05in}
\begin{equation}
\begin{aligned}
\label{CR: con1}
r_t = 0, \ \  d_t = R, \ \forall t \in [T_S, T_E],
\end{aligned}
\vspace{-0.05in}
\end{equation}
where $[T_S, T_E]$ is a subset of $[1, T]$, representing that only at some $t$ during a day, an ESS is used to provide contingency reserves. For the rest of the day, the ESS is not used. When providing contingency reserve, the ESS keeps discharging at the fixed rate as the reserve value $R$. In order to provide the maximal amount of reserves, an ESS is charged to its full energy capacity before response, i.e.,
\vspace{-0.05in}
\begin{equation}
\begin{aligned}
\label{CR: con2}
e_{T_S} = E_{cap}.
\end{aligned}
\vspace{-0.1in}
\end{equation}

We formulate the optimization problem for ESS in contingency reserves  by putting Eq.\eqref{gm: first}-\eqref{gm: last} together with Eq.\eqref{CR: credit}-\eqref{CR: con2}. The objective function is still to maximize the net profit.
% which equals to the revenue for providing reserves minus the amortized cost of ESS equipment. 
The decision variables are the same as those of RSR.

% Putting Eq.\eqref{gm: first} - Eq.\eqref{gm: last} together with Eq.\eqref{CR: credit} - Eq.\eqref{CR: con2}, the optimization formulation can be simplified as:
%\vspace{-0.05in}
%\begin{small}
 % \begin{equation}
  %\begin{aligned}
  %\label{CR:simple1}
%{\underset{E_{cap}, P_{cap}, R}{\max}} \ \ \  \Pi^{CR} R - (\Pi^{P,d} P_{cap} + \Pi^{E,d} E_{cap} ), \\
%\textbf{s.t.} \ \ \  (1-\mu)^T e_0 - \sum_{t=0}^{T-1}(1-\mu)^t R \geq (1 -  DoD) E_{cap},\\
%e_0 = E_{cap},\\
%R \leq P_{cap}, \\
%P_{cap} \geq 0, E_{cap} \geq 0, R \geq 0. \\
 % \end{aligned}
  %\vspace{-0.05in}
 % \end{equation}
%\end{small}
%\vspace{-0.2in}

\begin{figure}[tb]
        \centering
        \vspace{-0.1in}
        \includegraphics[width=0.5\columnwidth]{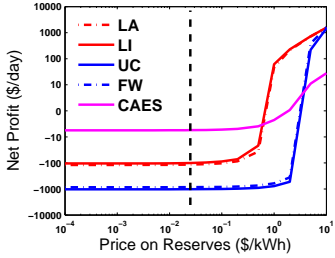}
        \vspace{-0.1in}
        \caption{The optimal net profit via varying contingency reserve prices $\Pi^{CR}$ for ESSs with today's typical capacities. The black dashed line shows where the current reserve price is at.}
         \label{fig: CR}
         \vspace{-0.2in}
  \end{figure}
  
%\vspace{-0.05in}
\subsubsection{Case Study}

We focus on the 10-minute spinning reserve as an example of contingency reserves, as it is expected to have the highest revenue. $\Pi^{CR}$ = \$0.025/kW is selected for the 10-minute spinning reserve based on today's market information~\cite{aikema2012data}. We assume the 10-minute spinning reserve is called once a day in our case, and $T_E - T_S = 10$min.

The optimal solution for all five ESSs in contingency reserve are: $P^*_{cap} = E^*_{cap} = R^* = 0$, which shows that none of five ESSs gain net profit by only providing contingency reserves at today's market reserve price, no matter what the power and energy capacities are used, and how they are operated. The larger the capacities ($E_{cap}$, $P_{cap}$) are used, the more reserves $R$ that an ESS can provide, however, as well as the higher the cost of ESS would be, and the cost is always larger than the revenue from providing $R$.

The $4^{th}$ row in Table~\ref{tb:compare} shows results of maximal net profit of contingency reserve and corresponding amount of reserve for today's typical ESS capacities, i.e., ($E_{cap}$, $P_{cap}$) given from Table~\ref{tb:typicalESD}. It highlights that none of today's typical ESSs earn profit from contingency reserves at today's reserve prices. Contingency reserves are demanding in terms of energy capacity (as opposed to power capacity), though the power capacity cannot be too low either. From the table, LA and LI batteries perform better than UC and FW, because of their lower price on energy capacity and relatively low self-discharge rate, but still not well enough to be profitable. %For example, CAES can only provide very limited amount of reserve due to the extremely small power density.
Fig.\ref{fig: CR} presents the optimal net profit via varying reserve prices $\Pi^{CR}$ for different ESSs. %The black dashed line shows the current market reserve price.
LI and LA batteries start to gain profit when the price is close to \$1/kWh, whereas the  critical points of CAES, UC and FW are around \$5-8/kWh.

\vspace{-0.1in}

\begin{figure}[tb]
\centering
\subfigure[1-day power trace.]{ 
%of a typical data center with HP workload.]{
\includegraphics[width=.22\textwidth]{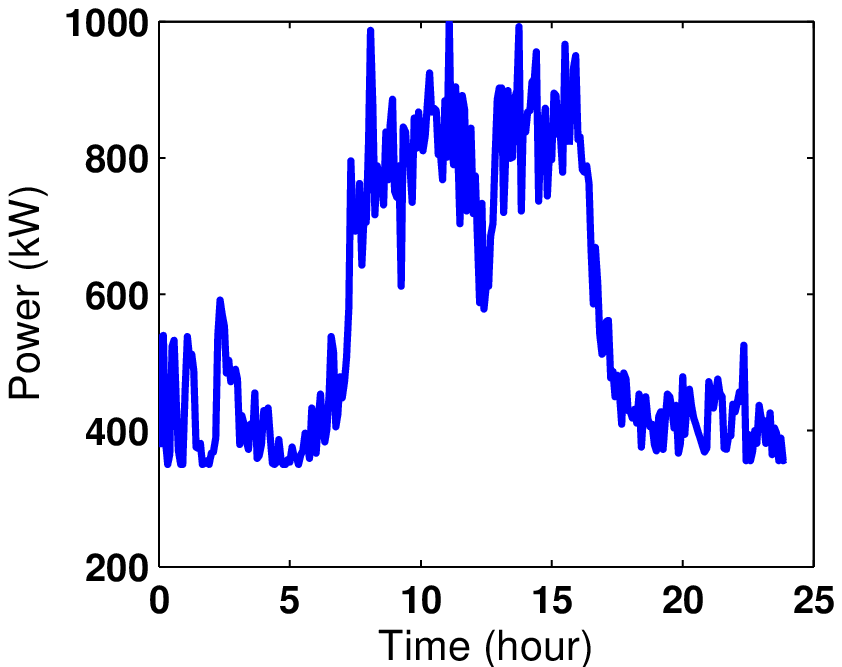}
 \label{fig:PowerCurve}
}\vspace{-0.1in}
\subfigure[Profit of LI batteries (\$/day). ]{
%via varying power and energy capacities.]{
\includegraphics[width=.22\textwidth]{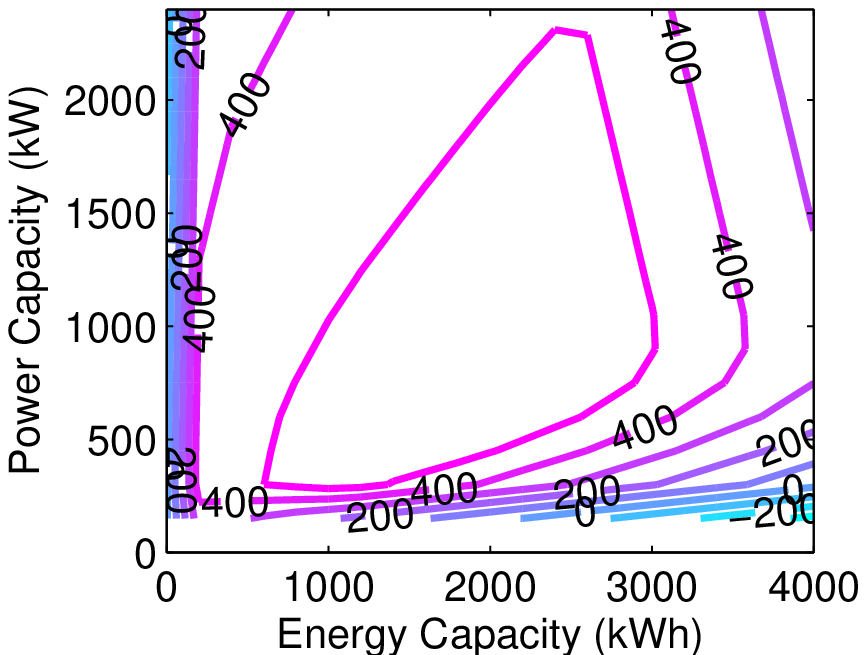}
 \label{fig:PeakS_Li_SAonPE_fval}
}
\subfigure[Profit of UC (\$/day). ]{
%via varying power and energy capacities.]{
\includegraphics[width=.22\textwidth]{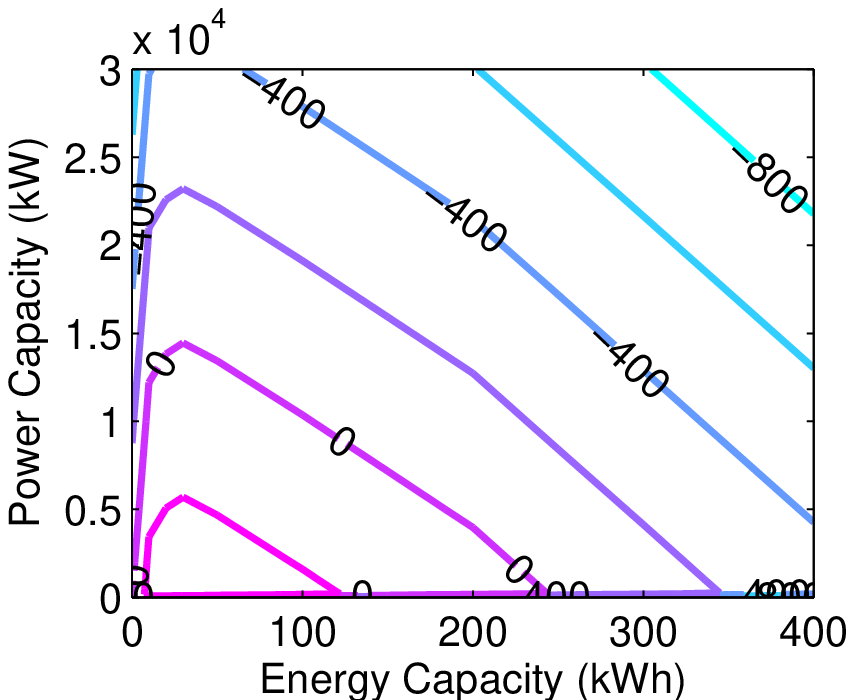}
\label{fig:PeakS_UC_SAonPE_fval}
}\vspace{-0.05in}
\subfigure[Profit of CAES (\$/day). ]{ 
%via varying power and energy capacities.]{
\includegraphics[width=.22\textwidth]{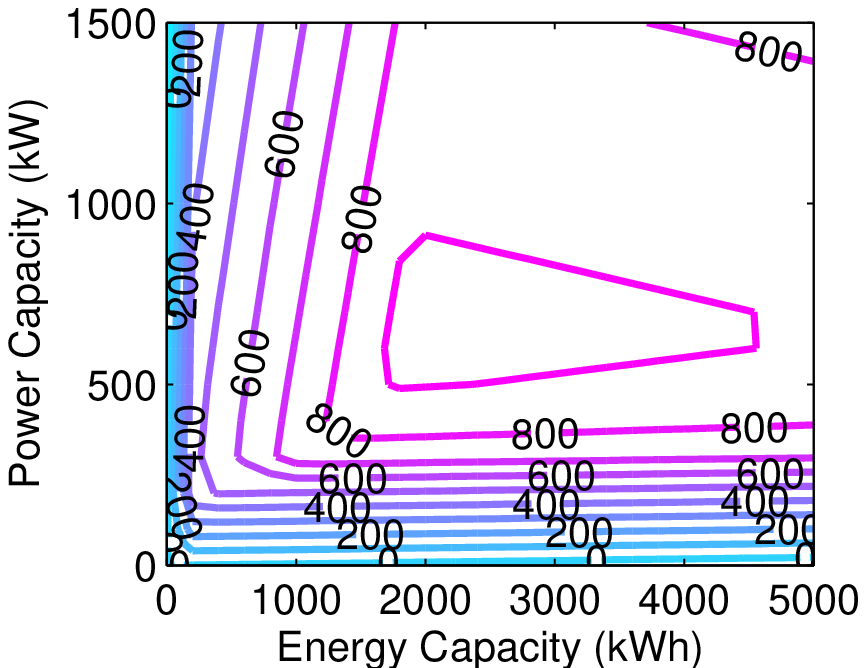}
\label{fig:PeakS_CAES_SAonPE_fval}
}
\subfigure[Impact of op-ex price on profit.]{
\includegraphics[width=.22\textwidth]{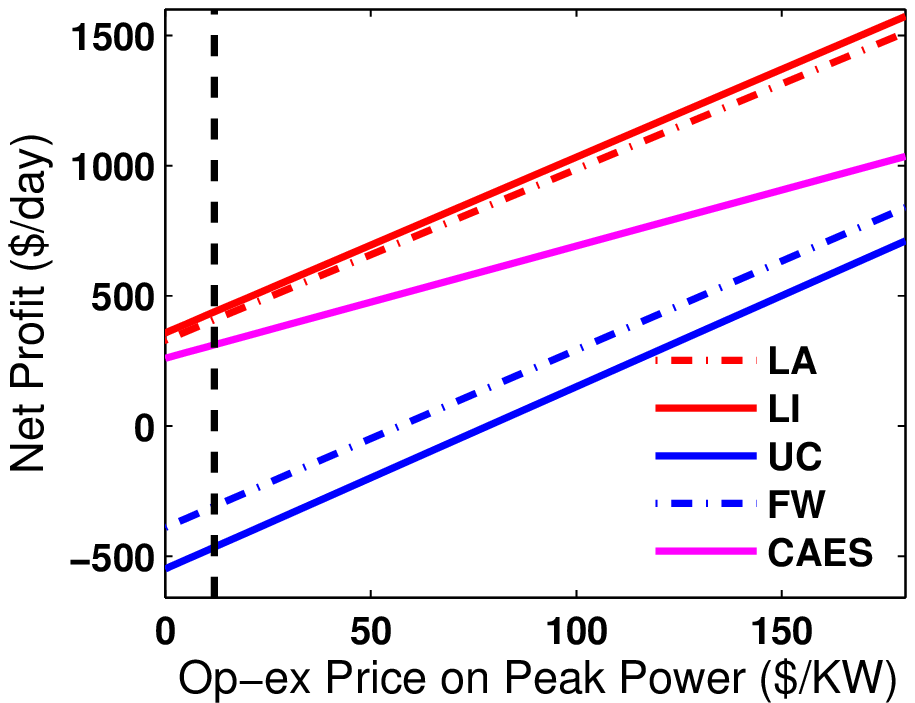}
\label{fig:PeakS_SAonOpexPrice}
}
\subfigure[Impact of cap-ex price on profit.]{
\includegraphics[width=.22\textwidth]{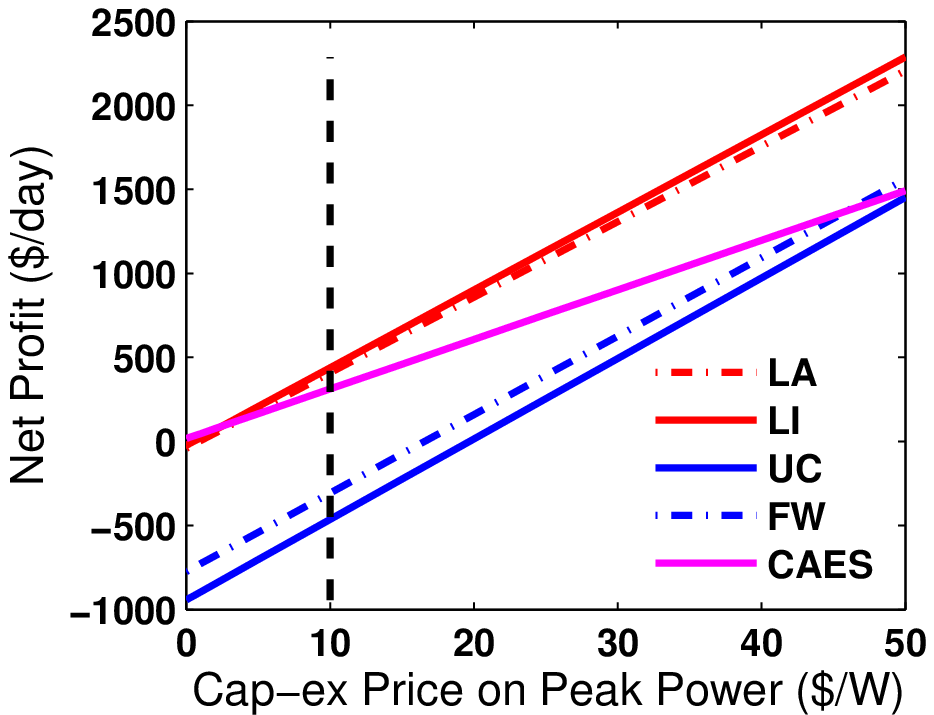}
\label{fig:PeakS_SAonCapexPrice}
}
\vspace{-0.1in}
\caption{ESSs in peak shaving. %(a) is an example of the daily power curve before peak shaving; (b) (c) and (d) are the optimal net profit via varying energy and power capacities ($E_{cap}$, $P_{cap}$), for LI, UC and CAES respectively; (d) and (f) are the optimal net profit via varying cap-ex and op-ex peak power prices respectively for multiple ESSs. The black dash lines show where the current market prices are around.
}
\label{fig:bin}
 \vspace{-0.2in}
\end{figure}

\subsection{Peak Shaving}
\label{sec: PS}
%\vspace{-0.05in}

The electricity bill charged monthly by utilities to large commercial and industrial power consumers, i.e., the operational expenditure (op-ex), typically consists of two parts: (i) the energy charge and (ii) the charge for the peak power during the month. The peak power is the maximum in the month of average power over each 15-30 minute duration. The price of the peak power (i.e., the op-ex peak power price) is around \$12/kW/Month currently. The one-time cost of building power infrastructure to provide capacities to satisfy the peak power requirements, i.e., the capital expenditure (cap-ex), is around \$10-20/W on peak power based on current estimates~\cite{PSUSigmetrics12}. Thus, cutting peak power is an important way to reduce costs. This approach, termed peak shaving, is common and ESS provides a key method for implementation.

 %\vspace{-0.05in}
\subsubsection{Problem Formulation}
When participating in peak shaving, an ESS that shaves $R$ amount of power from the peak power can gain revenue:
\vspace{-0.05in}
 \begin{equation}
  \begin{aligned}
  \label{eq: PSobj}
Revenue_{PS} = \Pi^{PS} R,
  \end{aligned}
  \vspace{-0.05in}
  \end{equation}
where $\Pi^{PS}$ is the overall price on shaved power, i.e., the summation of the amortized capital (cap-ex) price and operational (op-ex) peak power price. The peak shaving constraints in formulation, i.e., $Constraint_{PS}$ are:
\vspace{-0.05in}
  \begin{equation}
  \begin{aligned}
  \label{eq: PScons}
 0 \leq p_t + u_t \leq max (p_t)-R , \ \forall t \in [1, T], \\
 e_0 = e_T, \\
 \end{aligned}
 \vspace{-0.05in}
  \end{equation}
 where $p_t$ is the power curve before peak shaving, and $max(p_t)$ is the original peak power. $u_t$ is the power change rate from the view of system level. $p_t + u_t$ is the new power curve after peak shaving, and $max (p_t)-R$ is the new peak power. $e_0 = e_T$ represents that energy stored in ESS is kept the same at the beginning and in the end of the time frame (in our study $T$ = 1 day). We formulate the optimization problem for ESS in peak shaving by putting Eq.\eqref{gm: first}-\eqref{gm: last} together with Eq.\eqref{eq: PSobj}-\eqref{eq: PScons}. The objective goal is to maximize the net profit and the decision variables are the same as those of RSR.

\begin{table}[tb]
\caption{Optimal Solutions for Peak Shaving.}
\label{tb:optimalPS}
\begin{minipage}{8cm}
\def\arraystretch{1.5}\tabcolsep 0.8pt
\def\thefootnote{a}\footnotesize
 \centering
\begin{tabular}{|c|c|c|c|c|c|}
\hline
& LA & LI& UC & FW & CAES\\
\hline
$P_{cap}^*$ (kW) & $1.30*10^3$ & 769.19 & 148.39  & 147.85 & 645.36\\
\hline
$ E_{cap}^*$ (kWh) & $2.15*10^3$ & $2.40*10^3$ & 29.82 & 29.93 & $1.83*10^3$\\
\hline
Profit (\$/day)& 607.40 & 592.57 & 326.68 & 354.08 & 933.94 \\
\hline
$R^*$(kW) & 377.75 & 399.04 & 148.39 & 147.85 & 388.80 \\
\hline
\end{tabular}
%\footnotetext[1]{\scriptsize $\bar{D}$ and $\sigma_D$ are mean and typical deviation of performance degradation; $\bar{\epsilon}$ and $\sigma_{\epsilon}$ are mean and typical deviation of tracking error.}
\end{minipage}
 \vspace{-0.2in}
\end{table}

% \vspace{-0.05in}
\subsubsection{Case Study}

We generate $p_t$ from a real HP workload trace collected from a data center that consists of 5,000 servers. The peak power of this trace is 1MW, commonly seen in today's mid-size data center, and matches with the typical capacities of ESSs. Fig.\ref{fig:PowerCurve} is an example of $p_t$ in a day.

% For our experiments, the power trace $p_t$ in Eq.\eqref{eq: PScons} is constructed based on a real HP workload utilization trace collected between May 15th 2006 and June 18th 2006. More specifically, we assume a scenario where peak shaving is conducted on power consumption of a typical data center that consists of $N$ = 5000 servers, with each server has an idle power $P_{idle}$ = 70W and the maximal power $P_{max}$ = 200W~\footnote{The values are based on the measurement of an AMD Magny Cours (Opteron 6172) based server in our lab.}. The data center shares the HP workload utilization trace $w(t)$, and thus, its power trace $p_t = N[(P_{max} - P_{idle})w(t) + P_{idle}]$. The peak power of this trace is $NP_{max}$ = 1MW, which is common in today's data center, and also matches on magnitude with the capacities of typical ESS  listed in Table~\ref{tb:typicalESD}. Figure~\ref{fig:PowerCurve} shows an example of $p_t$ for a normal weekday. The peak power appears from 8am to 4pm, mainly in the working hours.

Unlike the optimal solution of RSR or contingency reserves that is either 0 or maximal capacity allowed (i.e., no feasible optimal solution), the optimal solution of peak shaving can be in between. Table~\ref{tb:optimalPS} lists the optimal solutions of different ESSs for peak shaving of the power trace $p_t$ shown in Fig.\ref{fig:PowerCurve}. All these optimal solutions lead to positive net profit. CAES has the maximal optimal net profit, though the corresponding capacities in the optimal solution is unrealistic due to its extremely small power and energy densities. LA and LI batteries have larger optimal net profit than UC and FW, though UC and FW can gain promising profit with very small capacities.

%\textcolor{red}{Then we study the sensitivity of net profit of peak shaving to ESS capacities.} 
Fig.\ref{fig:PeakS_Li_SAonPE_fval} to~\ref{fig:PeakS_CAES_SAonPE_fval} show the optimal net profit for varying energy and power capacities ($E_{cap}$, $P_{cap}$) in peak shaving, for LI, UC and CAES, respectively. These contour plots present where the optimal solution for each ESS is located. Fig.\ref{fig:PeakS_Li_SAonPE_fval} also shows that LI batteries can gain profit from peak shaving in most cases, except when the power capacity is very small. In Fig.\ref{fig:PeakS_UC_SAonPE_fval}, the profit of UC is larger than 0 only when both power and energy capacities are small, which shows that the marginal increase of the credit received from peak shaving by enlarging UC capacities is smaller than the increase in UC capacity cost. In Fig.\ref{fig:PeakS_CAES_SAonPE_fval}, CAES is always able to gain profit in peak shaving though large profit is not practical due to the limitations of power and energy densities.

Next, considering today's typical ESS capacities in peak shaving, the last row in Table~\ref{tb:compare} shows the optimal net profit and the corresponding optimal shaved power $R^*$ of ESSs with typical capacities in Table~\ref{tb:typicalESD}, and under today's cap-ex and op-ex market prices. From the table, UC and FW fail to gain net profit, whereas LA, LI and CAES earn net profit around \$300-400 per day.

Fig.\ref{fig:PeakS_SAonOpexPrice} and Fig.\ref{fig:PeakS_SAonCapexPrice} presents the optimal net profit of peak shaving for multiple ESSs, via varying op-ex and cap-ex peak power prices, respectively. The black dashed lines show where the current market prices are around. Note that in Fig.\ref{fig:PeakS_SAonOpexPrice}, the cap-ex price is fixed at \$10/W, while in Fig.\ref{fig:PeakS_SAonCapexPrice} the op-ex price is fixed at \$12/kW/Month (both of them are current prices). Fig.\ref{fig:PeakS_SAonOpexPrice} illustrates that CAES, LI, and LA gain net profit (larger than 0) under most cases including the current situation, while UC and FW need much higher payment to gain net profit. Similar results hold for cap-ex price in Fig.\ref{fig:PeakS_SAonCapexPrice}. %op-ex peak power price is fixed at \$12/kW/Month and the cap-ex price varies. CAES gains net profit no matter what the cap-ex price is. LA and LI batteries start to gain profit with a tiny cap-ex price, whereas UC and FW start to gain net profit when cap-ex price is around \$20/W. On the other hand in Figure~\ref{fig:PeakS_SAonOpexPrice}, cap-ex price is fixed at \$10/W and op-ex peak power price varies. CAES, LA and LI gain net profit no matter what the op-ex peak power price is, whereas UC and FW start to gain profit with the op-ex peak power price around \$60-80/kW/Month, which is 5-6 times of today's value.

%One thing we would like to point out here is,

The peak shaving results presented here can be generalized to any scenario as long as its power trace has a similar pattern to Fig.\ref{fig:PowerCurve}. This pattern is common in many scenarios~\cite{PSUSigmetrics12}, such as, weekday power consumption of offices, buildings and industries, power consumption of many types of data centers, e.g., data centers dealing with search workload (e.g., Google), communication workload (e.g., MSN), commercial and financial workload (e.g., stock exchange), etc. %In fact, this type of power trace is commonly seen in many scenarios, such as, weekday power consumption of offices, buildings and industries, power consumption of many types of data centers, including search workload (e.g., Google) based data centers, communication workload (e.g., MSN) based data centers, as well as commercial and financial workload (e.g., stock exchange) based data centers, etc. The peak shaving results on power traces with other patterns may be slightly different based on previous studies~\cite{PSUSigmetrics12}. For example, the power trace of a data center that mainly service high performance computing (HPC) workload is usually composed of several peak spikes distributed across the whole time frame. Such a trace has shorter bandwidth of each peak but higher frequencies of peak occurring. In that case an ESS with a stronger power capacity and more charge/discharge cycles may be preferred, and a strong energy capacity is no longer required.

\vspace{-0.1in}
\subsection{Discussion}
\label{sec: compare}
%\vspace{-0.05in}

\begin{table}[tb]
\caption{Comparing the Optimal Net Profit of Multiple Types of ESSs (with $E_{cap}$, $P_{cap}$ listed in Table~\ref{tb:typicalESD}) in Participating Different Programs. }
\label{tb:compare}
\begin{minipage}{8cm}
\def\arraystretch{1.5}\tabcolsep 1pt
\def\thefootnote{a}\footnotesize
 \centering
 \resizebox{8.5cm}{!} {
\begin{tabular}{|c||c|c||c|c||c|c||c|c||c|c||}
\hline
&  \multicolumn{2}{|c||}{LA} &  \multicolumn{2}{|c||}{LI} &  \multicolumn{2}{|c||}{UC} &  \multicolumn{2}{|c||}{FW} &  \multicolumn{2}{|c||}{CAES}\\
\hline
& \text{Profit} & $R^*$ & \text{Profit} & $R^*$ & \text{Profit} & $R^*$ & \text{Profit} & $R^*$ & \text{Profit} & $R^*$ \\
\hline
RSR & -16.4k  &0.17 & -11.1k & 0.29 & 13.0k &5.95& 10.3k &5.94& -0.3k& 0.004 \\
\hline
CR & -0.12k  &1.00& -0.10k &1.00& -1.02k &1.50& -0.85k &1.49& -0.006k& 0.02\\
\hline
PS & 0.41k &0.20& 0.44k &0.20& - 0.46k &0.21& -0.31k &0.20& 0.31k &  0.13\\
\hline
\end{tabular}
}
\footnotetext[1]{\scriptsize the unit of profit and $R^*$ in table are \$/day and MW.}
\footnotetext[2]{\scriptsize CR: contingency reserve; PS: peak shaving.}
\end{minipage}
\vspace{-0.2in}
\end{table}

% To summarize our illustration of the capabilities \textcolor{red}{and benefits} of ESS technologies across different demand response programs, 

We provide the optimal net profit of each ESS technology across the programs in Table~\ref{tb:compare} for today's typical capacities and market reserve prices. From the table, LA, LI batteries and CAES gain profit from peak shaving, whereas UC and FW gain profit from RSR. None of them gain profit from contingency reserve, due to its low price and low calling frequency. The maximal profit earned from emerging RSR (by today's typical UC or FW) is up to 30 times of the maximal profit that can be earned  from traditional peak shaving program (by LA or LI batteries), which shows that there is a great opportunity for an ESS to gain significant profit from RSR provision in today's ancillary market. For providing RSR, UC and FW are the best choices due to their extremely high tolerance for frequent charging/discharging, high efficiency and power density, and relatively low power capacity cost, while LA, LI batteries and CAES are better choices for peak shaving, or contingency reserves (though are not profitable), because of their relatively lower cost on energy capacity and lower self-discharge rate. 

% All the results show that there is a great opportunity for an ESS to gain profit by participating emerging RSR provision in today's ancillary market. In particular, RSR is more beneficial than other programs if a suitable type of ESS is chosen. \textcolor{red}{As an emerging market opportunity, RSR provision by ESS technology is still far from being well studied.} Thus, we focus the next section on designing policies to manage \textcolor{red}{ESS} participation in such programs. However, it is important to note that different types of ESSs earn maximal profit from different programs. So, a hybrid design including different types of ESSs for participating in multiple demand response programs may be beneficial.

% \vspace{-0.1in}

\section{Managing Participation in \\ Regulation Service Reserves}
\label{sec: policy}

%\textcolor{red}{Being an emerging market opportunity, RSR provision by ESS technology is still far from being well studied.} 
Given the potential profitability of ESS participation in RSR program, we now focus on the design of policies to enable this participation in practice. There are many challenges involved in such participation. For example, the provider is required to track an RSR signal that varies rapidly, bidirectionally, and is not known ahead of time.
% Complicated probabilistic constraints  are typically put on the tracking performance rather than simple deterministic constraints. 
In addition, the revenue is deducted by tracking error, which creates a trade-off between reserve maximizing and signal tracking. In this section, we start by developing offline optimal solutions (assuming the RSR signal is known a priori), 
%since the offline problem is already challenging in the general case, 
and then design practical online policies, in which the RSR signal is not known in advance. 
% Though online policies are heuristic and not as optimal as the offline policies because of the lack of information of RSR signal, it is more applicable for the real-life RSR participation.}
\vspace{-0.1in}
\subsection{Offline policies for RSR}
\label{sec: heuristicRS}
%\vspace{-0.05in}

In Section~\ref{sec: RSR}, we introduce the offline optimal solution in the case when $\rho_2 = 1$ in Eq.\eqref{RS: cons}. $\rho_2 = 1$ simplifies the probabilistic constraint in Eq.\eqref{RS: cons} to a deterministic constraint in Eq.\eqref{eq: 100Cons}. However, normally $\rho_2 < 1$ in practice, i.e., some violations of signal tracking are tolerable, which makes the optimization challenging.

\subsubsection{Policy overview}

In this section, we propose three heuristic offline solutions to deal with the probabilistic constraint in Eq.\eqref{RS: cons} when $\rho_2 < 1$. The key idea behind these solutions is to determine when the signal should be tracked within the tolerance $\rho_1$ (i.e., satisfying Eq.\eqref{eq: 100Cons}), and when the tolerance can be violated. Three solutions are as follows:
%\vspace{-0.05in}
% \begin{description}
%\item [RandSelect]

{\bf RandSelect}: Randomly select $\rho_2T$ time intervals in $[1, T]$ to satisfy Eq.~\eqref{eq: 100Cons}.

%\vspace{-0.05in}
%\item [MinCapSelect]
{\bf MinCapSelect}: Select $\rho_2T$ time intervals in $[1, T]$ with smallest $|\beta_t|$ to satisfy Eq.\eqref{eq: 100Cons}. This design is based on the fact that tracking RSR signal at the time interval $t$ with larger $|\beta_t|$ requires larger power capacity.

%\vspace{-0.05in}
%\item [FixIntSelect]
{\bf FixIntSelect}: Equally distribute $T-\rho_2T$ time intervals that are allowed to violate the Eq.\eqref{eq: 100Cons} in $[1,T]$. This is for the purpose of enabling the policy to adjust amount of energy stored in ESSs without closely following the tracking constraint once a while.
% \end{description}
%\vspace{-0.05in}

\subsubsection{Case study}

\begin{figure}[tb]
\centering
\subfigure[]{
\includegraphics[width=.22\textwidth]{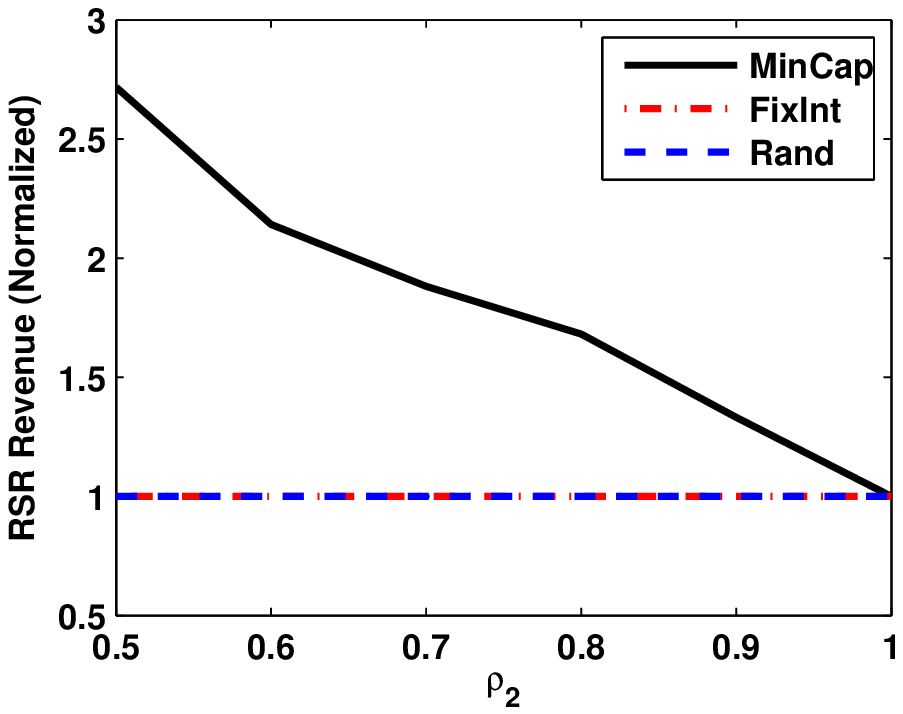}
 \label{fig:Li_rho2}
}
\subfigure[]{
\includegraphics[width=.22\textwidth]{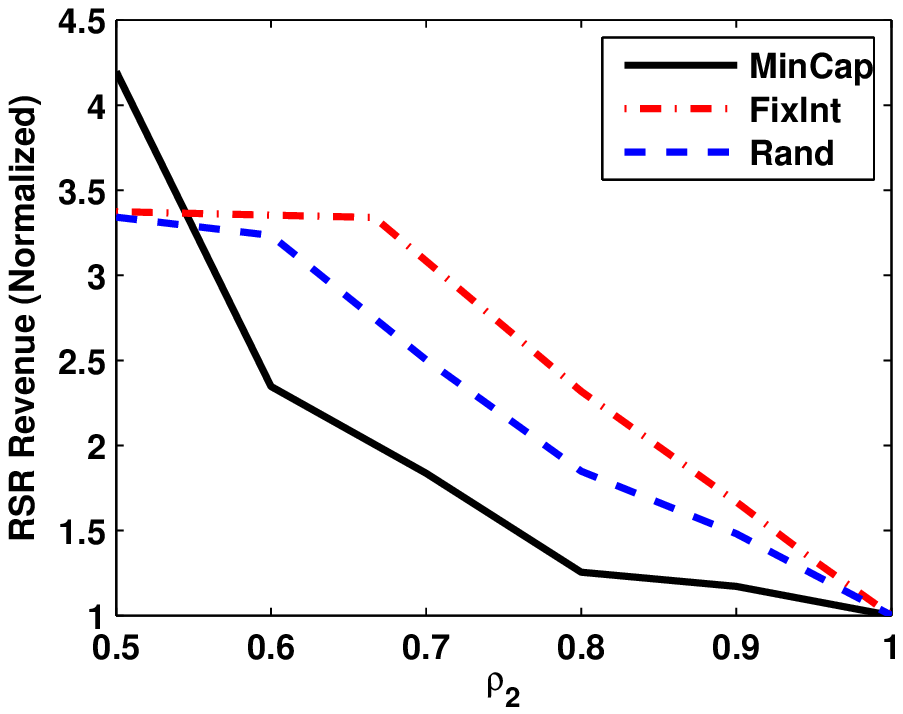}
 \label{fig:UC_rho2}
}
\vspace{-0.1in}
\caption{The revenue of providing RSR via varying $\rho_2$, for LI batteries (in~\ref{fig:Li_rho2}) and UC (in~\ref{fig:UC_rho2}), with three heuristic offline solutions, respectively. The revenue is normalized to the value of $\rho_2 = 1$.}
\label{fig:rho2}
 \vspace{-0.2in}
\end{figure}

Fig.\ref{fig:rho2} shows the optimal RSR revenue solved based on Eq.\eqref{RS:whole} with three proposed offline methods via varying $\rho_2$, for LI batteries and UC with typical capacities listed in Table~\ref{tb:typicalESD}, respectively. $\rho_1$ is fixed at 0.2, as in Section~\ref{sec: RSR}. Note that since we use the typical capacities in all cases, the cost of ESS is fixed. Thus, it is equivalent to make comparisons of these three methods based on either the RSR revenue, i.e., $Revenue_{RS}$ or the net profit originally used in the objective function of Eq.\eqref{RS:whole}. In the figure, all the revenues are normalized by the revenue at $\rho_2 =1$.

From Fig.\ref{fig:Li_rho2}, {\it MinCapSelect} always achieves largest revenue for LI batteries when $\rho_2$ varies. The charge/discharge capacities, i.e., the power capacity are the main bottleneck for LI batteries to offer more reserves, while {\it MinCapSelect} can help reduce the requirement on power capacity by only tracking small $|\beta_t|$ and giving up tracking large $|\beta_t|$, hence enabling LI batteries to provide additional reserves. The results for UC, however, are different. The power capacity is no longer the bottleneck, as today's typical UC has a much stronger power capacity compared to its energy capacity. As a consequence, energy capacity turns out to be the bottleneck. In that case, {\it MinCapSelect} does not help, and is even worse than the random algorithm {\it RandSelect}. A solution that is able to utilize the limited energy capacity in a more efficient way can provide more reserves 
%Thus, intervals that \textcolor{red}{are selected to be allowed to violate the tracking tolerance} become significant, as in those intervals, the amount of energy stored in ESS can be adjusted accordingly. 
and earn higher revenue. {\it FixIntSelect} becomes a better solution shown in Fig.\ref{fig:UC_rho2}, because it equally distributes time points where constraint violations are allowed across the whole time frame, so that the energy amount stored in ESS can be adjusted periodically and uniformly. Fig.\ref{fig:rho2} also shows that the optimal revenue increases when $\rho_2$ decreases. Relaxing the signal tracking constraints by decreasing $\rho_2$ in general offers more flexibilities for ESSs to participate the RSR program, and therefore, enables them to gain larger profits.

%\textcolor{red}{{\bf From Hao: I suggest to remove this following whole paragraph to save space. }}. Fig.\ref{fig:rho2} also demonstrates that results of UC are more sensitive to $\rho_2$ than those of LI batteries. For LI batteries, only {\it MinCapSelect} provides notable increases in revenue when the constraint is relaxed from $\rho_2=1$ to smaller values, whereas revenues of {\it RandSelect} and {\it FixIntSelect} do not change at all. However, all three heuristic solutions of UC have increases of revenue with smaller $\rho_2$. Relaxing the constraint by decreasing $\rho_2$ in general offers more flexibilities for an ESS to adjust its amount of energy stored, which is beneficial for those ESSs with bottleneck on energy capacity, such as UC and FW. Such a constraint relaxation, however, in general does not decrease the maximal power that an ESS is required to track\footnote{The maximal power to track decreases only when the relaxing time intervals are carefully chosen, e.g., in {\it MinCapSelect}.}. Thus it does not eliminate the requirement on power capacity, and has no major influence on results of ESSs with bottleneck on power capacity, such as LI  and LA batteries. In fact, it is expected that $\rho_1$ in Constraint~(\ref{RS: cons}) has greater effects on required power capacity than $\rho_2$. Therefore, it is expected that results of LI batteries are sensitive to $\rho_1$, while results of UC are not.

\vspace{-0.1in}
\subsection{Online policies for RSR}
\label{sec: online}

Prior offline solutions are based on the fact that RSR signal is known a priori, which is, however, not for the real case in practice. RSR signal is broadcast to demand side every few second in real time. In this section, we propose heuristic online ESS operational policies for RSR participation, where no information on the RSR signal is required in advance. In a practical scenario, the online policies handle the following problems: given the types and capacities of the ESS (i.e., assuming the ESS has been setup), how much reserve should be provided and how the ESS should be operated so that higher revenue from RSR participation can be gained and the feasibility of the participation can be guaranteed. 

%Note that such online policies are more applicable in practice, though may not be as optimal as the offline solution due to the lack of signal information. The studies of offline heuristics described in Section~\ref{sec: heuristicRS} provide insight into the design of efficient online policies. We propose online policies for different types of ESSs, and then compare their results with offline solutions.

\subsubsection{Policy overview}

As discussed before, {\it MinCapSelect} provides the highest revenue for ESSs such as LI and LA batteries in the offline solution. Hence we design the online operational policies for LI and LA batteries based on the {\it MinCapSelect} solution, as follows: 
%\begin{enumerate}
%\vspace{-0.05in}
%\item 

\textbf{Initialization:} we calculate two thresholds $\theta_0$ and $\theta_1$, based on the requirement input ($\rho_1$, $\rho_2$) from the market operator introduced before, and the historical data of RSR signal $\beta^{H}_t$, such that:
\vspace{-0.05in}
\begin{equation*}
\begin{aligned}
\textbf{Prob} \{ |\beta^{H}_t| \leq \theta_0 \} =  \rho_2, \\
\theta_1 =(1- \rho_1)\theta_0.
\end{aligned}
\end{equation*}
\vspace{-0.1in}
%\item 

\vspace{-0.05in}
\textbf{Real-time Operation:} at each time $t$, assuming the RSR signal value is $\beta^{RT}_t$, we determine the power rate $u_t$ by:
%\vspace{-0.05in}
\begin{enumerate}
\item If $|\beta^{RT}_t| < \theta_1$: we set $u_t = \beta^{RT}_t$, i.e., accurately track the signal;
\item If $\theta_0  \geq |\beta^{RT}_t| \geq \theta_1$: we set  $u_t =\theta_1 \textit{sign}(\beta^{RT}_t)$, i.e., cap the power rate $u_t$ at $\theta_1$;
\item If $|\beta^{RT}_t| > \theta_0$: we no-longer track the signal, instead, we set $u_t$ to adjust the current energy stored $e_t$ back to a middle level $e_m = \frac {DoD * E_{cap}} {2(1-\mu)}$ for future use (recall that $\mu$ is the self discharge rate);
\item Check and cap $u_t$ and $e_t$ based on power and energy capacity ($P_{cap}$, $E_{cap}$) constraints of the ESS.
\end{enumerate}
%\end{enumerate}
%\vspace{-0.05in}

An advanced algorithm could be updating $\theta_0$ and $\theta_1$ adaptively and dynamically in real time based on tracking performance feedback.

For ESSs such as UC and FW, the {\it FixIntSelect} solution offers the highest revenue from the previous study of the offline solution. Therefore, we propose the online operational policy for UC and FW based on the {\it FixIntSelect} heuristic, as follows:
% \vspace{-0.05in}

\textbf{Initialization:} we calculate the intervals that adjust the stored energy in ESS based on the input $\rho_2$: $T_{int} = \lceil \frac{1}{1-\rho_2} \rceil$, i.e., we adjust the stored energy every $T_{int}$ period. In addition, we set $\theta_1 =1- \rho_1$;
%\vspace{-0.05in}
% \item 

\textbf{Real-time Operation:} at each time $t$, assuming the RSR signal value is $\beta^{RT}_t$, we determine the power rate $u_t$ by:
\begin{enumerate}
%\vspace{-0.05in}
\item Every $t =T_{int}$, we set $u_t$ to adjust the current energy stored $e_t$ back to middle level $e_m = \frac {DoD * E_{cap}} {2(1-\mu)}$;
%\vspace{-0.1in}
\item For $t \neq T_{int}$, if $|\beta^{RT}_t| < \theta_1$: we set $u_t = \beta^{RT}_t$, i.e., accurately track the signal;
\item For $t \neq T_{int}$, if $|\beta^{RT}_t| \geq \theta_1$: we set  $u_t =\theta_1 \textit{sign}(\beta^{RT}_t)$, i.e., cap the power rate $u_t$ at $\theta_1$;
\item Check and cap $u_t$ and $e_t$ based on power and energy capacity ($P_{cap}$, $E_{cap}$) constraints of the ESS.
\end{enumerate}
%\end{enumerate}
%\vspace{-0.05in}

Another essential issue in an online policy is the determination of the amount of reserve to provide, i.e. $R_{onl}$. Unlike the offline solution, in which the RSR signal is known ahead, thus an optimal $R$ can be calculated directly from the optimization formulation, the $R_{onl}$ for the online policies is required to be carefully estimated. We propose an approach to learn $R_{onl}$ from historical offline solutions, as $R_{onl} = \lambda R_{min}$, where $R_{min}$ is the minimum of the offline optimal $R$ in the past 12 hours (the signal has been known in those hours, so offline optimal $R$ can be calculated), $\lambda$ is a discount value. We use $R_{min}$ and select $\lambda$ to avoid aggressive estimation of $R_{onl}$, 
%considering that an online solution can almost never be as good as the offline solution, 
and to guarantee feasibility of %$R_{onl}$ has the top priority in our case. 
our policies. We select $\lambda = 90\%$ for LI batteries and $\lambda = 75\%$ for UC, because LI batteries have more stable results, much smaller provision and are less sensitive to variations of $\rho_2$ than UC shown in Section~\ref{sec: heuristicRS}.

% We evaluate this solution by conducting the following preliminary study: We split our 24 RSR signals into training (first 12 hours) and testing (last 12 hours) sets. We learn $R_{onl}$ from the training set, as $R_{onl} = \lambda R_{min}$, where $R_{min}$ is the minimal offline $R$ among 12 training cases, $\lambda$ is a discount value. We use $R_{min}$ and select $\lambda$ to avoid the aggressive estimation of $R_{onl}$, considering that an online solution can almost never be as good as the offline solution, and guaranteeing the feasibility of $R_{onl}$ has the top priority in our case. We select $\lambda = 90\%$ for LI batteries and $\lambda = 75\%$ for UC, because LI batteries have more stable results, much smaller provision and less sensitive to variations of $\rho_2$ than UC shown in Section~\ref{sec: heuristicRS}. We then evaluate $R_{onl}$ to our 12 test cases to check whether its a feasible online solution for them. Moreover, we also evaluate $R_{onl}^{\rho_2}$ for different $\rho_2$. Our testing results show that these safely estimated $R_{onl}$ satisfy all test case constraints and are feasible solutions for all cases, for both LI batteries and UC.

\subsubsection{Case study}

An aggressive claim of $R_{onl}$ may lead to failure in reserve provisioning (i.e., constraints are violated) during the real-time operation, due to the limitations of ESS capacities. Hence, we first evaluate the feasibility of our online policies. We test the feasibility of our policies in the last 12 hours of a 1-day RSR signal. Each hour is a test case. In each test, we first calculate $R_{onl}$ based on the offline optimal $R$ in previous 12 hours as proposed, and then simulate the online policies to check whether all constraints are satisfied during the test hour. We also evaluate the policies with different $\rho_2$. Our results show that these safely estimated $R_{onl}$ together with our policies satisfy all constraints and thus are feasible solutions in all test cases, for both LI batteries and UC.

% We now move to an evaluation of the two online policies proposed above. The setting used is the same as described above. Results are shown in Fig.\ref{fig:rho2_off_on}. Revenue of offline solutions ({\it MinCapSelect} for LI batteries and {\it FixIntSelect} for UC) are also plotted in the figure for comparison. All results are normalized to the offline solution of $\rho_2 =1$.

\begin{figure}[tb]
\centering
\subfigure[]{
\includegraphics[width=.22\textwidth]{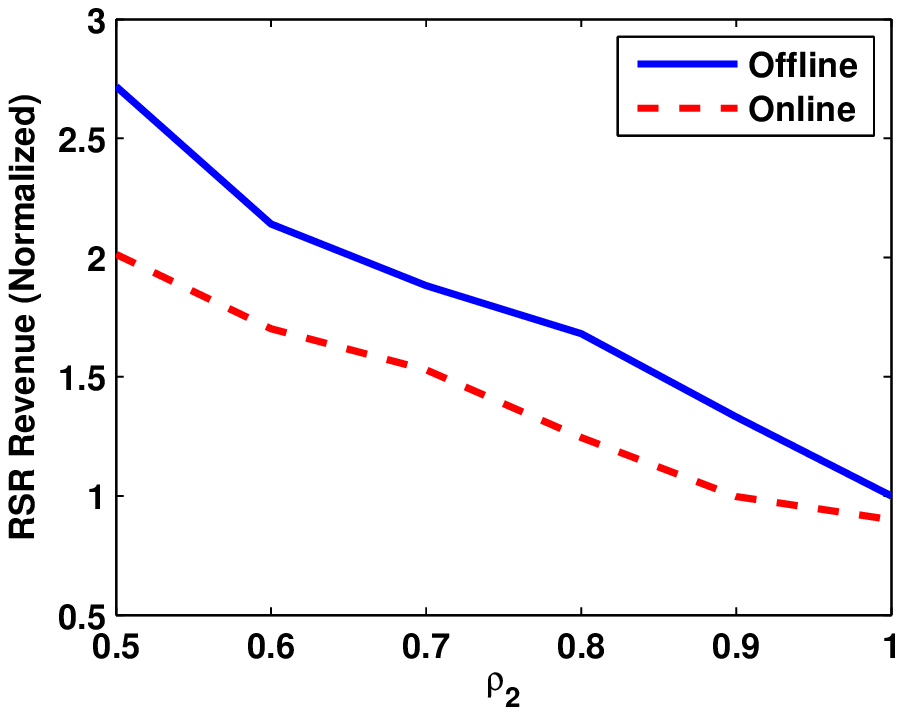}
 \label{fig:Li_rho2_offon}
}
\subfigure[]{
\includegraphics[width=.22\textwidth]{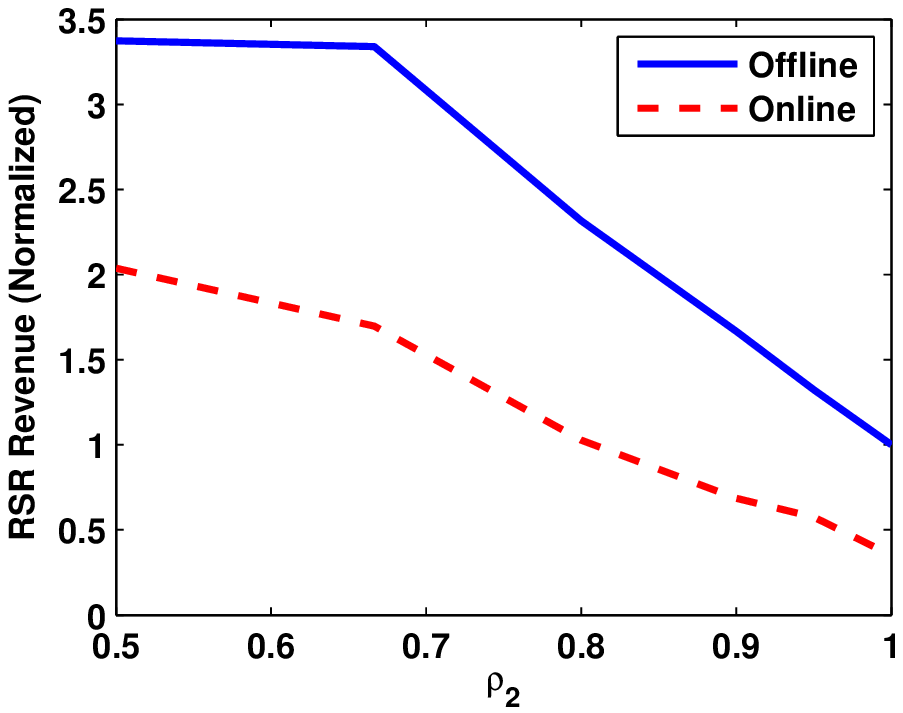}
 \label{fig:UC_rho2_off_on}
}
\vspace{-0.1in}
\caption{The revenue of providing RSR via varying $\rho_2$, for LI batteries (in~\ref{fig:Li_rho2_offon}) and UC (in~\ref{fig:UC_rho2_off_on}), respectively, with offline and online solutions. The revenue is normalized to the value of $\rho_2 = 1$ in offline solutions.}
\label{fig:rho2_off_on}
 \vspace{-0.2in}
\end{figure}

Then we compare the RSR revenue of our online policies to the offline solutions in Fig.\ref{fig:rho2_off_on}, via varying $\rho_2$. For offline solutions, {\it MinCapSelect} is selected for LI batteries, and {\it FixIntSelect} is selected for UC, as they perform the best for LI batteries and UC respectively shown in Fig.\ref{fig:rho2}, and our online policies are designed based on them. All results in Fig.\ref{fig:rho2_off_on} are normalized to the offline solution of $\rho_2 =1$. From the figure, the proposed online solutions still receive promising revenues, though there is (as expected) a noticeable gap compared to offline solutions, due to the lack of RSR signal information, and the safe estimation of the reserve value $R_{onl}$. More importantly, however, the feasibility of such online policies is guaranteed with high confidence. There is the following tradeoff: an aggressive online policy may bring the revenue close to optimal offline solutions, while the real-time feasibility of such solution decreases at the same time.

 %\vspace{-0.1in}
% \vspace{0.3in}
\section{Related Work}
\label{sec: related}

% Energy storage systems (ESSs) have advanced significantly and been widely studied in recent years. 
Today's most popular ESSs include batteries, flywheels, ultra-capacitors and other emerging techniques, e.g., CAES, etc~\cite{mccluer2008comparing, SmithAdavance}. These ESSs are modeled, for either ideal or non-ideal behaviors, and their system performance is evaluated~\cite{PSUSigmetrics12, ghiassi2013towards}. Recently, the hybrid electric energy storage system (HESS) is designed and investigated to enlarge the system storage capacity and improve the efficiency~\cite{pedram2010hybrid}.

%Ghiassi-Farrokhfal et al.~\cite{ghiassi2013towards} present an analytical ESS model for non-ideal ESS behavior, and investigate its effect on system performance. Different ESSs are usually designed for specific purposes and applied in very different scenarios \cite{SmithAdavance}. 
%Some recent work investigates hybrid electric energy storage system (HESS) designs that comprise heterogeneous EES elements to improve the system efficiency, enlarge the storage capacity, as well as to optimize the system cost~\cite{pedram2010hybrid}.

In tandem with the developments of ESSs, there is a growing attention on consumer (e.g., data centers, smart buildings and EVs) demand response and reserve provision in ancillary service markets.  A few studies closely explore opportunities and challenges in demand response and ancillary service market for data centers to reduce cost~\cite{liu2014pricing, chenIGCC, aikema2012data, WiermanIGCC, goiri2015matching}.
% Wierman et al.~\cite{WiermanIGCC} survey the opportunities and challenges of data center demand response. A few studies then closely explore and evaluate options in demand response and ancillary service market for data centers to reduce cost~\cite{liu2014pricing, chenIGCC, aikema2012data, goiri2015matching}.  
Among them, RSR is especially of interest due to its high clearing price on reserves, and thus potentially large profits~\cite{chenASPDAC}. Other work proposes to jointly leverage a data center and Plug-in Hybrid EVs in regulation market to maximize the profit~\cite{brocanelli2013joint}. Some approaches co-schedule heating, ventilation, air conditioning (HVAC) and EVs for reducing the energy consumption and the peak energy demand~\cite{wei2014co}.

ESSs are considered promising options for participation in power markets and demand response. A few previous studies propose control policies and evaluate the benefit of ESSs in real-time dynamic energy pricing programs~\cite{wang2013optimal, zhu2013maximizing}, peak shaving~\cite{PSUSigmetrics12, aksanli2013architecting}, and frequency control~\cite{4282047, cho2013enhanced}, respectively. However, most previous studies focus on traditional power market programs, though, ESSs are able to potentially receive higher profit from emerging ancillary service market, especially from RSR. In the space of RSR, some prior work surveys potential market chances and evaluates maturity of ESS participation in RSR~\cite{walawalkar2007economics, kumaraswamy2013evaluating, vu2009benefits, 7007628}, but without formulating the detailed models of participation and evaluating the optimal solutions. The closest paper to the current work is~\cite{YoungIGCC14}. However,~\cite{YoungIGCC14} uses a simplified RSR participation model that does not consider the details of regulation accuracy constraints and penalties. Further, it assumes that the RSR signal always follows a statistical distribution known a priori, and without considering the reserve value and capacity planning for different ESSs. To the best of our knowledge, ours is the first paper to provide detailed models, evaluate and optimize the profits of various ESS technologies in not only traditional power market programs such as peak shaving, but also in emerging smart grid demand response such as RSR and contingency reserves, by proposing detailed reserve value and capacity planning, as well as online ESS operational policies.  
 %\vspace{-0.2in}
%\vspace{-0.2in}
\section{Conclusion}
\label{sec: conclusion}

In this paper, we have modeled and studied the optimization solutions that maximize the net profit of various ESSs in different demand response programs. 
%We have evaluated the ESS technology most appropriate for each program and, vice versa, the most profitable program for each ESS technology. 
Our results show that typical UC and FW are the most profitable selections for RSR, while common battery techniques such as LI and LA batteries are the best choices for peak shaving. None of today's ESS technologies can earn positive net profits from merely providing contingency reserves. More importantly, applying UC/FW in RSR has the potential to be up to 30 times more profitable than LI/LA batteries for peak shaving. Additionally, we have proposed online policies for managing ESS participation in RSR program, the novel but most profitable option according to our studies. Our online policies guarantee the feasibility of RSR provisions, while also achieving significant profits. 
%In future work, we plan to evaluate the feasibility and benefit of providing reserves by jointly leveraging ESSs and other demand response techniques, e.g., data center power regulation and cooling control. Other important directions include developing a hybrid design to manage multiple ESSs  participating multiple programs. 
%\vspace{-0.05in}

% use section* for acknowledgement
\section*{Acknowledgment}
This paper is supported by the NSF Grant 1464388.

%The authors would like to thank...

\bibliographystyle{unsrt}
{\scriptsize
\bibliography{hao} % sigproc.bib is the name of the Bibliography in this case
}

% conference papers do not normally have an appendix

% trigger a \newpage just before the given reference
% number - used to balance the columns on the last page
% adjust value as needed - may need to be readjusted if
% the document is modified later
%\IEEEtriggeratref{8}
% The "triggered" command can be changed if desired:
%\IEEEtriggercmd{\enlargethispage{-5in}}

% references section

% can use a bibliography generated by BibTeX as a .bbl file
% BibTeX documentation can be easily obtained at:
% http://www.ctan.org/tex-archive/biblio/bibtex/contrib/doc/
% The IEEEtran BibTeX style support page is at:
% http://www.michaelshell.org/tex/ieeetran/bibtex/
%\bibliographystyle{IEEEtran}
% argument is your BibTeX string definitions and bibliography database(s)
%\bibliography{IEEEabrv,../bib/paper}
%
% <OR> manually copy in the resultant .bbl file
% set second argument of \begin to the number of references
% (used to reserve space for the reference number labels box)
%\begin{thebibliography}{1}

%\bibitem{IEEEhowto:kopka}
%H.~Kopka and P.~W. Daly, \emph{A Guide to \LaTeX}, 3rd~ed.\hskip 1em plus
%  0.5em minus 0.4em\relax Harlow, England: Addison-Wesley, 1999.

%\end{thebibliography}

% that's all folks
\end{document}